\documentclass[fleqn,usenatbib]{mnras}

\usepackage{newtxtext,newtxmath}
\usepackage{amsmath}
\usepackage{booktabs}
\usepackage{tabularx}
\usepackage{multirow}

\usepackage[T1]{fontenc}
\usepackage{subcaption}
\DeclareRobustCommand{\VAN}[3]{#2}
\let\VANthebibliography\thebibliography
\def\thebibliography{\DeclareRobustCommand{\VAN}[3]{##3}\VANthebibliography}

\usepackage{graphicx}	
\usepackage{amsmath}	
\usepackage{float}
\usepackage{cuted}
\usepackage{xcolor}
\usepackage[normalem]{ulem} 

\title[Non-linear velocities in the stacked kSZ]{Interpreting the stacked kinetic SZ effect I: velocity reconstruction and non-linear velocity effects}

\author[Ondaro-Mallea et al.]{
Lurdes Ondaro-Mallea$^{1,2}$\thanks{E-mail: lurdes.ondaro@ast.cam.ac.uk},
Raul E. Angulo$^{3,4}$,
Boryana Hadzhiyska$^{1,2}$, 
Joop Schaye$^{5}$\\
$^{1}$Institute of Astronomy, University of Cambridge, Madingley Road, Cambridge CB3 0HA, United Kingdom\\
$^{2}$Kavli Institute for Cosmology Cambridge, Madingley Road, Cambridge CB3 0HA, UK\\
$^{3}$ {Donostia International Physics Center, Manuel Lardizabal Ibilbidea, 4, 20018 Donostia, Gipuzkoa, Spain}\\
$^{4}$IKERBASQUE, Basque Foundation for Science, 48013, Bilbao, Spain \\
$^{5}$ Leiden Observatory, Leiden University, PO Box 9513, 2300 RA Leiden, the Netherlands\\
}

\date{Accepted XXX. Received YYY; in original form ZZZ}

\pubyear{2015}

\begin{document}
\label{firstpage}
\pagerange{\pageref{firstpage}--\pageref{lastpage}}
\maketitle

\begin{abstract}
The stacked kinetic Sunyaev–Zel'dovich (kSZ) signal probes the velocity-weighted projected gas momentum around galaxies, and is emerging as a powerful probe of gas fractions and baryonic feedback. Its interpretation, however, rests on several assumptions that we test in this pair of companion papers. Using the FLAMINGO hydrodynamical simulations and DESI-like galaxy mocks for luminous red galaxies (LRGs), the bright galaxy sample (BGS), and emission-line galaxies (ELGs), we identify the ingredients required to model the signal to better than $10\%$. This first paper focuses on velocities. We decompose the signal into a dominant bulk-flow term, proportional to the mean optical depth, plus non-linear terms arising from the small-scale gas momentum and its coupling to the stacking velocity. When the stacking velocities are reconstructed from linear information in real space alone -- an idealisation which is not possible in practice -- the non-linear terms cancel and the signal traces the mean optical depth to within a few per cent. When the stacking velocity instead retains non-linear information or is affected by redshift-space distortions, the non-linear terms suppress the signal by $10-20\%$: a $1-2\sigma$ effect for current data that is expected to be statistically significant for upcoming surveys, and one that depends only weakly on baryonic feedback. Our results reveal a trade-off: velocity estimators that retain small-scale information boost signal-to-noise but require simulation-based modelling, whereas conservative reconstructions simplify the interpretation at the cost of signal-to-noise.
\end{abstract}





\section{Introduction}

Current and next-generation surveys are mapping the Universe with unprecedented precision. Weak-lensing surveys, in particular, provide increasingly tight constraints on the matter distribution \citep{Abbott2016DES, deJong2015KiDS, Iye2004Subaru, EuclidCollaboration2022, Ivezic2019LSST}, with much of their constraining power coming from small-scale clustering. In these scales, baryonic physics and other systematic effects become important \citep[e.g.][]{vanDaalen:2011}. Robustly modelling and constraining these effects is therefore essential to fully exploit the cosmological information contained in these surveys.

While baryons follow the dark matter distribution on large scales, they are substantially redistributed within and around galaxies and haloes by poorly understood astrophysical processes, including stellar feedback and feedback from \textit{active galactic nuclei} (AGN). Predicting these effects from first principles in cosmological volumes remains beyond the reach of current simulations. Instead, hydrodynamical simulations rely on effective sub-grid models, whose different calibrations can lead to markedly different predictions for the matter distribution and its clustering \citep[e.g.][]{Semboloni:2011, Chisari:2019, vanDaalen:2020}. 


An alternative is to constrain the gas distribution around haloes directly through observations. The traditional route has been to measure ``halo gas fractions'' from the X-ray emission of the hot intracluster medium \citep[e.g.][]{Sun:2009, Eckert:2016}. These observations tend to probe the high-mass end of the halo population and point to relatively mild feedback scenarios, in which no more than roughly half of the associated gas mass has been evacuated from haloes of $M \sim 10^{14}\,h^{-1}\,\rm M_{\odot}$ \citep[e.g.][]{Grandis:2024}.

In recent years, the kinetic Sunyaev--Zel'dovich (kSZ) effect \citep{Sunyaev:1972,Sunyaev:1980} -- the Doppler shift imprinted on cosmic microwave background (CMB) photons when they scatter off free electrons in moving haloes -- has emerged as a competitive probe of gas in the halo outskirts. The combination of high-resolution CMB experiments \citep{Thornton2016ACTPol} with large galaxy surveys \citep{Dawson2013BOSS, DESI:2016} has enabled high signal-to-noise stacked kSZ measurements \citep{Schaan:2021, Hadzhiyska:2024, RiedGuachalla:2025, Hadzhiyska:2025, Roper:2025, Hadzhiyska:2026, Qu:2026}. These are obtained by stacking CMB temperature maps at the positions of a large galaxy sample, weighting each by an estimate of the galaxy's line-of-sight velocity \citep{Li:2014}. This velocity is typically reconstructed by solving the linearised continuity equation on the galaxy density field \citep{Hadzhiyska:2023, RiedGuachalla:2023}. Current measurements already reach $>10\sigma$ statistical significance, and with next-generation CMB experiments \citep{Ade2019SO} they are expected to exceed $30\sigma$ \citep{so:2025}.

The first analyses of these measurements have inferred halo gas fractions noticeably lower than those favoured by previous X-ray observations \citep[e.g.][]{Sun:2009, Eckert:2016} or by most state-of-the-art hydrodynamical simulations \citep[e.g.][]{Schaye:2015, McCarthy:2017, Pillepich:2018, Dave:2019, Parkmor:2023, Schaye:2023}, with several recent works converging on this conclusion \citep{Hadzhiyska:2024, Bigwood:2024, McCarthy:2025a, Kovac:2025, Roper:2025, Siegel:2025, Siegel:2026}.

As these measurements become increasingly precise, the theoretical models used to interpret them must improve accordingly. 
For upcoming surveys, theoretical uncertainties of only a few per cent will already be comparable to the statistical uncertainties \citep{so:2025}.

Interpreting stacked kSZ measurements requires understanding which physical processes shape the observed signal and which modelling ingredients are required for unbiased constraints on baryonic feedback and halo gas fractions. In this series of papers, we address these questions using realistic mock observations. By isolating individual physical effects, we quantify their impact on stacked kSZ profiles and assess the validity of the approximations commonly adopted in their interpretation. This first paper focuses on the role of velocity reconstruction and non-linear velocity effects, while Paper~II will investigate the impact of satellite galaxies and its degeneracy with baryonic physics.

Formally, the low-redshift stacked kSZ signal measures the velocity-weighted projected gas momentum around the galaxy sample. Because peculiar velocities are coherent over large scales, the stacked signal is expected, to leading order, to trace the projected gas density, with the velocity field providing only an overall normalisation. In this paper, we test the accuracy of this approximation and determine the regimes in which additional velocity-dependent contributions become important.

Recent work has shown that this simple picture may break down for various reasons. Correlations between the density and velocity fields can make a non-negligible contribution to the stacked kSZ signal \citep{Wayland:2026}, while baryonic feedback alters gas velocities over large scales and may therefore also affect the measured profile \citep{OndaroMallea:2025}. Furthermore, inaccuracies in the reconstructed velocity field can modify the inferred kSZ profile in ways that extend beyond a simple rescaling of its amplitude \citep{RiedGuachalla:2023,Qu:2026}.

In this paper we derive all mathematically allowed contributions to the stacked kSZ estimator and determine which are non-zero and which are relevant at the precision of current and forthcoming measurements. We use the FLAMINGO suite of cosmological hydrodynamical simulations \citep{Schaye:2023,Kugel:2023} which provides three key ingredients: i) realistic mock galaxy samples whose masses and clustering properties closely match those used in observational analyses; ii) multiple feedback models, allowing us to isolate the impact of baryonic feedback on both the density and velocity fields; iii) fully non-linear gravitational evolution, which naturally couples these fields. We perform the analysis both in an idealised setup, where the true line-of-sight velocities are known, and in a realistic setup, where they are reconstructed self-consistently from the mock galaxy catalogue in the same way as in observations.

We find that the standard picture remains accurate when velocities are reconstructed using conventional linear methods with sufficiently large smoothing scales. In this regime, the stacked kSZ signal traces the projected gas density to within $\approx  5\%$. The approximation breaks down, however, once the reconstructed velocity contains small-scale information or is affected by redshift-space distortions, producing corrections at the $10-20\%$ level.

The paper is organised as follows. Section~\ref{sec:decomposition_theory} decomposes the stacked kSZ estimator into its possible contributions 
(summarised in Table~\ref{tab:ksz_decomposition}). Section~\ref{sec:measurements} describes the numerical setup and the methods used to compute the mock profiles. Section~\ref{sec:results} presents the results both for true line-of-sight velocities and for reconstructed ones. Section~\ref{sec:implications} discusses the implications for observational strategies and theoretical modelling. Section~\ref{sec:conclusions} summarises and concludes.

\section{Decomposition of the stacked kSZ signal}\label{sec:decomposition_theory}

The kinematic Sunyaev-Zel'dovich (kSZ) effect arises from the Doppler shift of CMB photons when they Thomson scatter off free electrons moving with a non-zero velocity relative to the CMB rest frame. The resulting temperature fluctuation is proportional to the electron column density weighted by their line-of-sight velocity. At low redshifts, where the intergalactic medium is highly ionized, the electron density is given by the gas density up to fixed normalizing factors. The kSZ signal measured around a galaxy is therefore proportional to the projected line-of-sight gas momentum,\begin{equation}\label{eq:ksz_effect}
    \frac{\Delta \mathcal{T}_{\rm kSZ}}{T_{\rm CMB}}(r_{\rm p}) \propto \int_{z=0}^{z_{\rm CMB}} \rho(\chi, r_{\rm p}) \, v(\chi, r_{\rm p}) \, d\chi,
\end{equation}
where $\rho$ is the gas density, $v$ is its line-of-sight velocity, $\chi$ is the line-of-sight coordinate, and $r_{\rm p}$ is the projected distance to the galaxy, with $\chi=0$, $r_{\rm p}=0$ denoting the centre of the galaxy. Here and throughout, we drop overall constants and sign conventions that do not affect the discussion.

Because the kSZ effect is intrinsically weak, high significance detections require stacking techniques \citep{Li:2014} \citep[the kSZ signal of galaxy clusters can be detected for individual objects, e.g.][]{Adam:2017}. Since galaxies are equally likely to be moving towards or away from the observer, a simple unweighted average over galaxy positions would vanish. To extract the signal, one instead performs a velocity-weighted average:
\begin{equation}\label{eq:ksz_estimator}
    T_{\rm ksz}(r_{\rm p}) = \frac{1}{r{v_{\rm rms}^{\rm stack}}}\left \langle \Delta \mathcal{T}_{\rm kSZ}(r_{\rm p}) \, v_{\rm stack} \right \rangle,
\end{equation}
where $v_{\rm stack}$ is the line-of-sight velocity assigned to each galaxy, $v_{\rm rms}^{\rm stack}$ is their root-mean-square (rms), and $r$ is the cross-correlation coefficient between stacking and true velocities, given by 
\begin{equation}
    r = \frac{\langle v_0 v_{\rm stack}\rangle}{v_{\rm rms}^0 v_{\rm rms}^{\rm stack}},
\end{equation}
where $v_{0}$ is the true bulk flow of gas and $v_{\rm rms}^0$ its rms. In observational stacks the velocities are not known and have to be estimated, while in simulations the true velocities are directly accessible.

The velocity-weighted stacked kSZ estimator therefore probes the mean, velocity-weighted, projected gas momentum field of a given galaxy sample. This receives contributions from the gas density field, the gas velocity field, the correlations between them, and their joint correlation with the stacking velocity.

We can further expand this expression by decomposing the gas velocity as
\begin{equation}\label{eq:norm_vel}
    v(\chi,r_{\rm p}) = v_{0} + v^\prime(\chi,r_{\rm p}), 
\end{equation}
where $v^\prime(\chi, r_{\rm p})$ is the deviation of gas velocities from the bulk flow of gas $v_0 = v(0,0)$ along the line of sight or across the projected radius. Note that we evaluate this equation per galaxy, and thus $v_0$ is different for each galaxy.
Substituting into Eq.~\ref{eq:ksz_estimator} yields
\begin{equation}\label{eq:ksz_halo1}
\begin{aligned}
    {T_{\rm ksz}}(r_{\rm p}) = \frac{1}{r\,v_{\rm rms}^{\rm stack}} \bigg[
        & \left\langle \underbrace{\left[\int d\chi\,\rho(\chi, r_{\rm p})\right]}_{\tau(r_{\rm p})} v_{0}\, v_{\rm stack} \right\rangle \\
        &{}+ \left\langle \underbrace{\left[\int d\chi\,\rho(\chi, r_{\rm p})\, v^{\prime}(\chi, r_{\rm p})\right]}_{p^{\prime}(r_{\rm p})} v_{\rm stack} \right\rangle \bigg] ,
\end{aligned}
\end{equation}
where $\tau$ is the projected gas mass\footnote{At low redshift, the mean optical depth and the mean projected gas mass differ only by a constant normalization factor. We therefore use the notation $\tau$ and refer to the two interchangeably.}, and $p^{\prime}$ is the projected gas momentum, arising from the decorrelation of the gas motion from the bulk flow at the galaxy position. Splitting both terms following $\langle XY\rangle = \langle X\rangle \langle Y \rangle  + \mathrm{Cov}(X,Y)$, and taking into account that $\langle v_{\rm stack}\rangle=0$,
\begin{equation}\label{eq:ksz_halo_split}
\begin{aligned}
    {T_{\rm ksz}}(r_{\rm p}) = \frac{1}{r\,v_{\rm rms}^{\rm stack}} \bigg[ & \left\langle \tau(r_{\rm p}) \right\rangle \left\langle v_{0}\, v_{\rm stack} \right\rangle \\
    &{}+ \mathrm{Cov}\!\left(\tau(r_{\rm p}),\, v_{0} v_{\rm stack}\right) \\
    &{}+ \mathrm{Cov}\!\left( p^{\prime}(r_{\rm p}),\, v_{\rm stack}\right) \bigg].
\end{aligned}
\end{equation}
Introducing the expression of the cross-correlation coefficient into the previous equation:
\begin{equation}\label{eq:ksz_halo2}
\begin{aligned}
    {T_{\rm ksz}}(r_{\rm p}) = {} & \underbrace{\left\langle \tau(r_{\rm p}) \right\rangle v_{\rm rms}^{0}}_{\text{mean bulk flow}} \\
    &+ \frac{1}{r\,v_{\rm rms}^{\rm stack}} \bigg[\,\underbrace{\mathrm{Cov}\!\left(\tau(r_{\rm p}),\, v_{0} v_{\rm stack}\right)}_{\text{density-bulk flow correlation}} \\
    &{}+ \underbrace{\mathrm{Cov}\!\left( p^{\prime}(r_{\rm p}),\, v_{\rm stack}\right)}_{\text{velocity decorrelation}} \bigg].
\end{aligned}
\end{equation}

These are the three mathematically possible contributions to the velocity-weighted kSZ signal, summarised in Table \ref{tab:ksz_decomposition}. Each term is sourced by a different process: 
\begin{itemize}
    \item \textit{Mean bulk flow component}: The first term is the signal sourced by the coherent motion of the gas along the line-of-sight. Provided the cross-correlation coefficient $r$ and $v_{\rm rms}^0$ are well characterized, it directly probes the mean optical depth of the galaxy sample. This makes it extremely useful for studying the baryonic effects around the sample. This term is purely physical: all information about the velocity estimation is encoded in $r$.
    \item \textit{Density-bulk flow correlation component}: The second term describes the small-scale correlation between the bulk flow of the gas and/or the stacking velocity with the gas density around the galaxy. This  can have contributions due to physical correlations ($\tau$ and $v_0$) and correlations induced by the velocity reconstruction ($\tau$ and $v_{\rm stack}$), provided both correlate ($v_0v_{\rm stack}$).
    \item \textit{Velocity decorrelation component}: The third term captures the correlation of the stacking velocity with the deviation of the gas momentum from the bulk flow $\propto \tau v_0$. Unlike the previous term, this arises purely from correlations between the small-scale gas momentum and the stacking velocity.
\end{itemize}
In linear theory, velocities possess very long coherence lengths and can therefore be assumed to be constant on the scales over which the density $\rho$ varies rapidly. Density and velocity are then effectively decoupled: the former is sourced primarily by small-scale fluctuations, the latter by large-scale ones. As a result, on scales $r<50 \, h^{-1}\,\rm Mpc$, $\mathrm{Cov}(\tau, v_{0} v_{\rm stack}) \approx 0$ and $v' \approx 0$. Thus, the density-bulk flow correlation and velocity decorrelation terms disappear.

However, beyond linear theory these terms can be non-zero. This can happen for two reasons: non-linear gravity, or errors in the halo velocity estimation that are correlated with non-linearities. We refer to these as the ``non-linear'' terms.

Finally, we apply a compensated aperture filter (CAP), which at each projected radial distance $r_{\rm p}$ takes the signal enclosed within $r_{\rm p}$ and subtracts an estimate of the background, computed as the signal in an adjacent equal-area annulus:
\begin{equation}
    T_{\rm ksz}^{\rm CAP}(<r_{\rm p}) = T_{\rm ksz}(<r_{\rm p}) - 
    T_{\rm ksz}(r_{\rm p}<r<\sqrt{2}\,r_{\rm p}).
\end{equation}
We apply the CAP filter to the total signal and to each component in the decomposition. This filter is extremely useful for mitigating the noise from primary CMB fluctuations in measurements. When applied to theoretical estimations, it removes the background contributions that are uncorrelated with the galaxy. For instance, while $\tau$ would have contributions from all the gas from $z=0$ to $z_{\rm CMB}$, $\tau^{\rm CAP}$ only picks up the gas density in the vicinity of the galaxy \citep{Hadzhiyska:2023b}. 

The goal of this paper is to quantify the relevance of the non-linear velocity terms in Eq.~\ref{eq:ksz_halo2} with realistic galaxy samples, and identify their physical origin. We do so in an idealised scenario where the stacking velocities are known exactly, and in a more realistic scenario where they are estimated via linear reconstruction from the galaxy density field, both with and without redshift-space distortions. In the next Section we describe the simulations, galaxy selections, and velocity estimators that we use to do so.

\begin{table}
\centering
\caption{Decomposition of the velocity-weighted stacked kSZ signal. The last two terms vanish under the linear-theory assumption for velocities, and we therefore refer to them as ``non-linear terms''. Although each can individually reach $\approx 50\%$ of the total kSZ signal, the net non-linear contribution is $\approx  10-20\%$.}
\begin{tabularx}{\columnwidth}{@{}p{3.2cm} X@{}}
\toprule
\textbf{Component} & \textbf{Description} \\
\midrule
Mean bulk flow
  & Signal sourced by the coherent, large-scale motion of the gas. Probes the mean optical depth times the RMS of the gas bulk flow. Dominant term, and independent of the stacking velocity. \\
\addlinespace
Density-bulk flow correlation
  & Signal sourced by the correlation between the small-scale gas density field around the galaxy, and the bulk flow times the stacking velocity ($v_0 v_{\rm stack}$). It can reach $\approx 50\%$ of the total kSZ signal at large projected distances. \\
\addlinespace
Velocity decorrelation
  & Signal sourced by the deviation of the gas velocity from the bulk flow ($v-v_0$) and its coupling to the stacking velocity ($v_{\rm stack}$). It can reach $\approx 50\%$ of the total kSZ signal at large projected distances, and $\approx 10\%$ on small scales when the sample contains satellite galaxies. \\
\bottomrule
\end{tabularx}
\label{tab:ksz_decomposition}
\end{table}

\section{Methods}\label{sec:measurements}
In this Section we describe the simulations (Section \ref{sec:sims}), galaxy selections (Section \ref{sec:galaxy_selection}), and velocity estimators (Section \ref{sec:stack_th}) that we use to construct mock kSZ signals (Section \ref{sec:estimator}).

\subsection{Simulations}\label{sec:sims}

We carry out our study using the FLAMINGO suite of cosmological hydrodynamical simulations, a state-of-the-art simulation set that includes several physics variants run in cosmological volumes \citep{Schaye:2023, Kugel:2023,Helly:2026}. In this work we use the $L=1000\,\rm Mpc$ box at the fiducial baryonic mass resolution of $m_{\rm p} = 1.07 \times 10^9 \, \rm M_{\odot}$. Unless stated otherwise, we show results for the $f_{\rm gas}-8\sigma$ physics variant, which is calibrated to reproduce suppressed gas fractions in clusters and is favoured by several recent studies (e.g. \cite{McCarthy:2025b}, but see \cite{Seppi:2026}). As we show in Section~\ref{sec:non_linear_bar}, our conclusions are dominated by gravitational dynamics rather than by hydrodynamic or astrophysical processes, and therefore generalise to any physics variant.

Haloes are identified using a three-dimensional FoF algorithm \citep[e.g.][]{Press:1982} with a linking length equal to 0.2 times the mean inter-particle separation. Subhaloes are then identified within the FoF groups using the \texttt{HBT-HERONS} algorithm, which uses information from the substructure history \citep{Han:2018,Forouhar:2025}. We have also repeated the analysis using \texttt{VelociRaptor} as the subhalo finder \citep{Elahi:2019}. While all qualitative conclusions are robust to the choice of subhalo finder, we find some subdominant quantitative differences. We attribute these primarily to differences in the classification of subhaloes as centrals or satellites.

We then use halo and galaxy properties computed with Spherical Overdensity $\&$ Aperture Processor \citep{McGibbon:2025}. Throughout this paper we define the halo radius $R_{\rm 200m}$ as the radius enclosing a mean density equal to 200 times the mean density of the universe, and the host halo mass $M_{\rm 200m}$ as the mass enclosed within $R_{\rm 200m}$.

\subsection{Galaxy samples}\label{sec:galaxy_selection}
Our main analysis uses a DESI luminous red galaxies (LRG)-like sample \citep{Zhou:2023}. We construct it by rank-ordering galaxies in the simulation at $z=0.7$ by stellar mass and selecting the $N$ most massive ones such that their mean halo mass is $\langle M \rangle = 10^{13.2}\, h^{-1} M_{\odot}$, as measured by stacking the CMB lensing signal in the same sample in \cite{Hadzhiyska:2025}. The resulting satellite fraction is $f_{\rm sat} = 0.23$. Satellites dominate the high-mass end of the selection, with mean host halo mass $ \log_{10} \langle M_{\rm sat}\rangle = 13.64$ versus $\log_{10}\langle M_{\rm cen}\rangle= 12.84$ for centrals, with minimum stellar mass of $\log_{10} M_*>10.85$, where all masses are in units of $h^{-1} \rm M_{\odot}$.

In Section~\ref{sec:implications}, we extend our analysis to two additional galaxy samples. The first is a DESI Bright Galaxy Survey (BGS)-like sample at $z=0.25$ \citep{Hahn:2023}. We select galaxies by stellar mass to match the mean halo mass inferred from CMB lensing, $\langle M \rangle = 10^{13.23}\,h^{-1}M_{\odot}$ \citep{Hadzhiyska:2026}. This yields a satellite fraction of $f_{\rm sat}=0.31$, $\log_{10} \langle M_{\rm sat}\rangle = 13.7$ versus $\log_{10}\langle M_{\rm cen}\rangle= 12.2$, and minimum stellar mass of $\log_{10} M_*>9.8$. 

The second is a DESI Emission Line Galaxy (ELG)-like sample at $z=1.1$ \citep{Raichoor:2023}. We select galaxies by star formation rate to match the mean halo mass inferred by \cite{Garcia-Quintero:2025} ($\langle M \rangle = 10^{12.2}\,h^{-1}M_{\odot}$; see also \citealt{Ortega:2026}). However, the FLAMINGO simulations do not have sufficient resolution to resolve galaxies at this mass. We therefore adopt the lowest halo mass that can be robustly sampled (with star formation rate larger than 0), yielding $\langle M \rangle = 10^{12.4}\,h^{-1}M_{\odot}$, with $\log_{10} \langle M_{\rm sat}\rangle = 12.81$ versus $\log_{10}\langle M_{\rm cen}\rangle= 11.93$, satellite fraction of $f_{\rm sat}=0.30$, and $\text{SFR} > 0.2667 \, h^{-1} \rm M_{\odot} / yr$.
 
We emphasise that, although our mock samples are realistic, we do not fit them to real samples; we therefore make no direct comparison between the measured and mock kSZ signals, comparing only with the fractional errors in Section~\ref{sec:implications}.

\subsection{Stacking velocity}\label{sec:stack_th}

Stacking the kSZ signal requires weighting the signal at each galaxy by the line-of-sight gas velocity at its position ($v_0$ in Eq.~\ref{eq:norm_vel}). In simulations, we have access to the true velocity (Sec. \ref{sec:stack_real_th}). In observations, the gas motion is not directly accessible and must be reconstructed from the observed galaxy density field. Here, we mimic the widely used method of linear reconstruction (Sec. \ref{sec:stack_vrec_th}).

\subsubsection{True velocity}\label{sec:stack_real_th}

For the true velocity, we use the line-of-sight halo velocity, $v_{\rm stack} = v_0 = v_h$. The halo velocity is defined as the centre-of-mass velocity of all particles within the halo. We have verified that using the centre-of-mass velocity of the gas particles instead produces no appreciable difference in the stacked signal, so we adopt halo velocities throughout for simplicity and generality.

For satellite galaxies, we use the velocity of their host halo rather than that of the subhalo or satellite itself \citep[see also][]{RiedGuachalla:2023}. Satellite galaxies can have substantially larger velocities than their hosts because of their orbital motion within the halo. However, the ionized gas surrounding satellites is largely stripped and therefore does not follow this orbital motion \citep{He:2026,Contreras:2026}. Assigning the satellite or subhalo velocity to the gas would therefore be unphysical. In practice, doing so increases the amplitude of the first term in Eq.~\ref{eq:ksz_halo2}, since $v_{\rm rms}^{\rm sub} \gg v_{\rm rms}^{\rm h}$, and artificially generates a large velocity decorrelation contribution (i.e.\ a large negative $v'(0,0)$ in Eq.~\ref{eq:norm_vel}). This makes the interpretation of the measurements more complicated. Beyond this internal effect on the decomposition of the signal, we find that using subhalo velocities also modifies the shape of the measured total kSZ profile at the $\approx20\%$ level on large scales (see Appendix \ref{sec:app_stack_vel}). As discussed later in the paper, this arises because of different coupling of satellite and halo velocities to the underlying gas momentum field.

In this case, the cross-correlation coefficient between true and stacking velocities is $r=1$ by definition. 

\subsubsection{Reconstructed velocity}\label{sec:stack_vrec_th}

The line-of-sight velocity of galaxies, $v_{\rm stack}$, is not known \textit{a priori} and must be estimated from observations. The standard approach is to compute the smoothed galaxy overdensity field, transform it into Fourier space, and solve the linearised continuity equation \citep[e.g.][]{Schaan:2021,RiedGuachalla:2023,Hadzhiyska:2023},
\begin{equation}\label{eq:vrec}
    \hat{\mathbf{v}}_{\rm rec} = -i\,a f H\,\frac{\mathbf{k}}{k^2}\,\frac{\hat{\delta}_g}{b+f\mu^2},
\end{equation}
where $\hat{\mathbf{v}}_{\rm rec}$ is the reconstructed velocity in Fourier space, $\hat{\delta}_g$ is the smoothed galaxy overdensity field in Fourier space, $\mathbf{k}$ is the wavevector, $a$ is the expansion factor of the universe, $f$ is the linear growth rate, and $H$ is the Hubble parameter. The linear bias of the galaxy sample, $b$, converts the galaxy overdensity to matter overdensity, and $f \mu^2$ corrects for redshift-space distortions to linear order, where $\mu$ is line-of-sight angle, $\mu=\mathbf{\hat{n}}\cdot \mathbf{k}/k$. The line-of-sight component of $\hat{\mathbf{v}}_{\rm rec}$ is then used as the stacking velocity.

We compute the reconstructed velocities from the simulated galaxy catalogues, mimicking observational procedures. To this end, we use the publicly available \texttt{pyrecon} package\footnote{\url{https://github.com/cosmodesi/pyrecon}}, running the reconstruction algorithm on the mock catalogues. We build the galaxy density field on a mesh with $n_{\rm grid} = 128$ (cell size $L/n_{\rm grid} = 5.32 \, h^{-1}\rm Mpc$) and further smooth it on a scale $R_s = 10\,h^{-1}\rm Mpc \approx 15 \, \rm Mpc$. In Section \ref{sec:implications}, we examined the impact of the smoothing scale by adopting $R_s = 30 \,h^{-1}\mathrm{Mpc}$. For comparison, current observational analyses use smoothing scales of $R_s = 12.5\,h^{-1}\mathrm{Mpc}$ \citep{Hadzhiyska:2026} and $R_s = 15\,h^{-1}\mathrm{Mpc}$ \citep{Qu:2026}.

We compute the bias of each galaxy sample as the ratio of the galaxy-matter cross-power spectrum to the matter auto-power spectrum, averaged over the range $0.01 < k\, [h/\mathrm{Mpc}] < 0.1$. The Hubble parameter and linear growth rate are evaluated for the FLAMINGO cosmology at the redshift of each galaxy sample.

We do the reconstruction both in real and redshift space. In real space, we build $\hat{\delta}_g$ from true galaxy positions, and adopt $f \mu^2 =0$ in the reconstruction. The corresponding cross-correlation coefficient with true halo velocities for the DESI LRG-like sample is $r\approx 0.8$.

We then distort the galaxy positions by their peculiar velocities, and build $\delta_g$ directly in redshift space. In this case, the cross-correlation coefficient with the true halo velocities is $r \approx 0.7$ for the DESI-like LRG and BGS samples, comparable to the values quoted by \cite{RiedGuachalla:2023}. 

For ELGs, we find a cross-correlation coefficient of $r\approx0.74$, somewhat higher than the value reported by \cite{Hadzhiyska:2026} ($r\approx0.55$), which was estimated using Abacus HOD catalogs. However, among the three galaxy samples considered, ELGs are arguably the least well resolved in FLAMINGO, and their HOD modelling is also the most uncertain. We therefore defer a thorough investigation of the origin of this discrepancy to future work.

\subsection{Estimator}\label{sec:estimator}

We compute the kSZ signal and its decomposition (Eq. \ref{eq:ksz_halo2}) around each galaxy in the sample out to $r_{\rm p} = 10\,h^{-1}\rm Mpc$ in the transverse plane, by projecting the gas momentum and mass within a $20\,h^{-1}\rm Mpc$-thick shell along the line of sight. For this we use the publicly available \texttt{halotools} package\footnote{\url{https://github.com/astropy/halotools}} \citep{Hearin:2017} with some modifications. We have checked that this configuration yields converged results to projecting the full box, both for the total kSZ signal and for each term of the decomposition. 

The position of each galaxy is taken to be the position of the most bound particle of its host halo (for central galaxies) or subhalo (for satellite galaxies). The bulk gas velocity required for the decomposition (Eq.~\ref{eq:norm_vel}) is taken to be the centre-of-mass velocity of the host halo, $v_0 = v_h$, as motivated in Section~\ref{sec:stack_real_th}. Note that this choice does not affect the total kSZ signal $\Delta \mathcal{T}_{\rm ksz}$ associated with each galaxy, but it can affect the relative importance of the terms in its decomposition (see Section~\ref{sec:cen_sat}).

We compute the stacked kSZ signal of the full sample, stacking with both true and reconstructed halo velocities (Section~\ref{sec:stack_th}), and apply the CAP filters as explained in Section \ref{sec:decomposition_theory}. We drop the explicit CAP label for brevity, but throughout this paper all profiles and discussion refer to CAP-filtered profiles.
\begin{figure*}
    \centering
    \includegraphics[width=\linewidth]{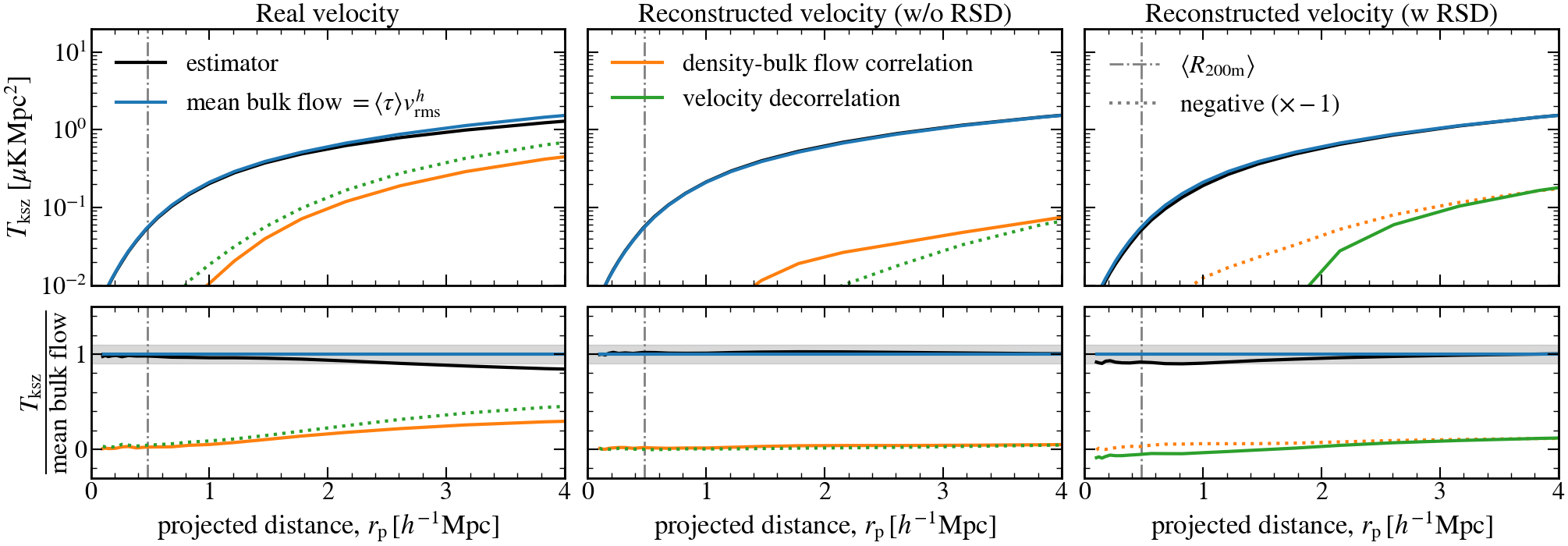}
    \caption{Decomposition of the velocity-weighted stacked kSZ signal for the full DESI-like LRG galaxy sample. Columns show results for different stacking velocities: the true halo velocity (left), the reconstructed velocity in real space (middle), and the reconstructed velocity in redshift space including RSD (right). Upper panels show profiles as a function of projected radius, while lower panels show each component normalized by the mean bulk flow contribution -- $T_{\rm ksz} / \langle \tau \rangle v_{\rm rms}^h$. The total signal is shown in black, the mean bulk flow contribution in blue, the density-bulk flow correlation term in orange, and the velocity decorrelation term in green. Vertical dash-dotted lines indicate the mean virial radius of the sample, and dotted curves denote quantities plotted with reversed sign. The mean bulk flow is the dominant contributor to the stacked kSZ signal, accounting for $\approx 80 - 90\%$ of the total signal across all velocity definitions. The combined effect of the non-linear terms is to suppress the signal, with a scale dependence set by the stacking velocity.}
    \label{fig:ksz_corr_decorr}
\end{figure*}

Since the goal of this paper is to provide general guidelines for modelling stacked kSZ signals, we minimise survey- and sample-specific choices, while keeping the highest possible realism. In particular, we display all profiles in comoving megaparsecs rather than arcminutes --- the latter requiring a specific redshift to do the conversion --- and we do not apply beam smoothing. The only exception is in Figure \ref{fig:scale_dependence_hvel_vrec}, where we apply the beam to be able to compare the results to observational error bars.

\section{Results}\label{sec:results}
In this section, we evaluate the relative importance of different terms in the decomposition of the stacked kSZ signal. These include the mean bulk flow component and the non-linear terms (Eq.~\ref{eq:ksz_halo2}, Section~\ref{sec:decomposition_theory}).

We consider three scenarios of increasing realism. First, we stack using the true halo velocities (Section~\ref{sec:stack_real}). Next, we use velocities reconstructed from galaxy catalogues using linear theory in real space (Section~\ref{sec:stack_vrec_real}). Finally, we include reconstructed velocities in redshift space, accounting for redshift-space distortions (Section~\ref{sec:stack_vrec_z}). This progression helps isolate the intrinsecally non-linear processes in the gas velocity field, and clarifies how they couple to the reconstructed velocities. Finally, we study how baryonic physics modifies these results (Section~\ref{sec:non_linear_bar}).

\begin{figure*}
    \includegraphics[width=0.9\linewidth]{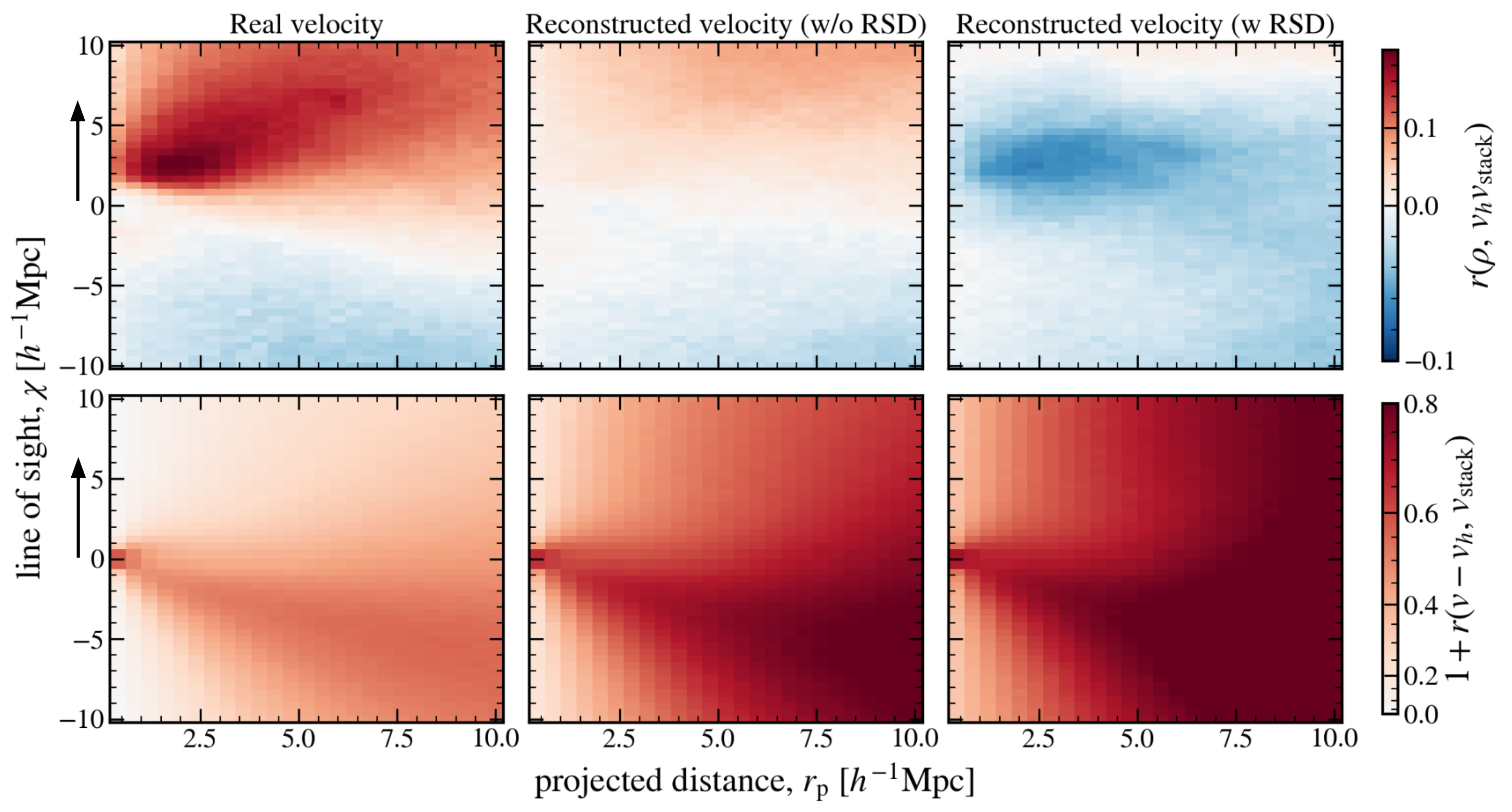}
    \caption{Correlation coefficient $r$ between the stacking velocity and the gas density (top row) or gas velocity (bottom row) around halos of DESI-like LRGs. Halos are located at the origin of each panel and move in the positive line-of-sight direction (arrows), where massive structures lie on average. Red indicates positive correlation, blue negative, and white no correlation. The top row shows the correlation with the environmental gas density, $r(\rho, v_h v_{\rm stack})$. The bottom row shows the decorrelation of the gas velocity from the halo velocity, $1 + r(v - v_h, v_{\rm stack})$. Red indicates gas moving with the halo, and white indicates full decorrelation. Columns correspond to three choices of stacking velocity: the true halo velocity (left), the reconstructed velocity without redshift-space distortions (RSD; middle), and the reconstructed velocity including RSD (right). Using the true velocity reveals the intrinsically non-linear structure of the gas velocity field. Massive structures in the environment boost the halo velocity (upper left panel) and reduce velocity coherence (lower left panel) in the direction of motion. In real space, reconstructed velocities do not inherit non-linear information. As a result, reconstructed velocities do not correlate with the small-scale density (upper middle panel), and the gas velocity can then be treated as coherent with the halo over a larger region (lower middle panel). Redshift-space distortions introduce reconstruction errors that correlate with the environment density. This produces an anti-correlation between density and reconstructed velocity (upper right panel). It also reduces velocity decorrelation in the direction of motion of the halo (lower right panel).}
    \label{fig:correlation_halovel_dens}
\end{figure*}

Three figures support this section, where we show the results for our fiducial DESI LRG-like sample. 
\begin{itemize}
    \item Figure~\ref{fig:ksz_corr_decorr} presents the decomposition of the stacked kSZ signal into its three components: the mean bulk flow (blue), the density--bulk flow correlation term (orange), and the velocity decorrelation term (green).
    
    \item Figure~\ref{fig:ksz_ratios_cen_sat} shows the same decomposition separately for central and satellite galaxies, allowing us to identify the origin of the different behaviours.
    
    \item Figure~\ref{fig:correlation_halovel_dens} illustrates the physical origin of the non-linear terms by showing how the gas density and velocity fields around the haloes in the sample correlate with the stacking velocity.
\end{itemize}
All three figures share the same layout. From left to right, the columns correspond to stacking with the true halo velocity, reconstructed velocities in real space, and reconstructed velocities in redshift space. The following subsections discuss these three cases in turn.

\subsection{Stacking with halo velocities}\label{sec:stack_real}

We begin with an idealised case in which the true halo velocities are used for stacking. This isolates the intrinsic non-linear contributions to the kSZ signal, without the additional complications introduced by velocity reconstruction.

Figure~\ref{fig:ksz_corr_decorr} (left column) shows that the stacked kSZ signal is dominated by the mean bulk flow component, which accounts for $80-90\%$ of the total signal. The remaining non-linear contribution reaches $10-20\%$ on large scales. 
Interestingly, the two non-linear terms are each much larger than their sum. Individually, they contribute up to $\approx50\%$ of the total kSZ signal, but with opposite signs, so they largely cancel. 

We first explain the physical origin of these two non-linear terms (Section~\ref{sec:non_linear_physics}), before examining why their relative importance differs for central and satellite galaxies (Section~\ref{sec:cen_sat}).

\begin{figure}
\centering
    \includegraphics[width=\linewidth]{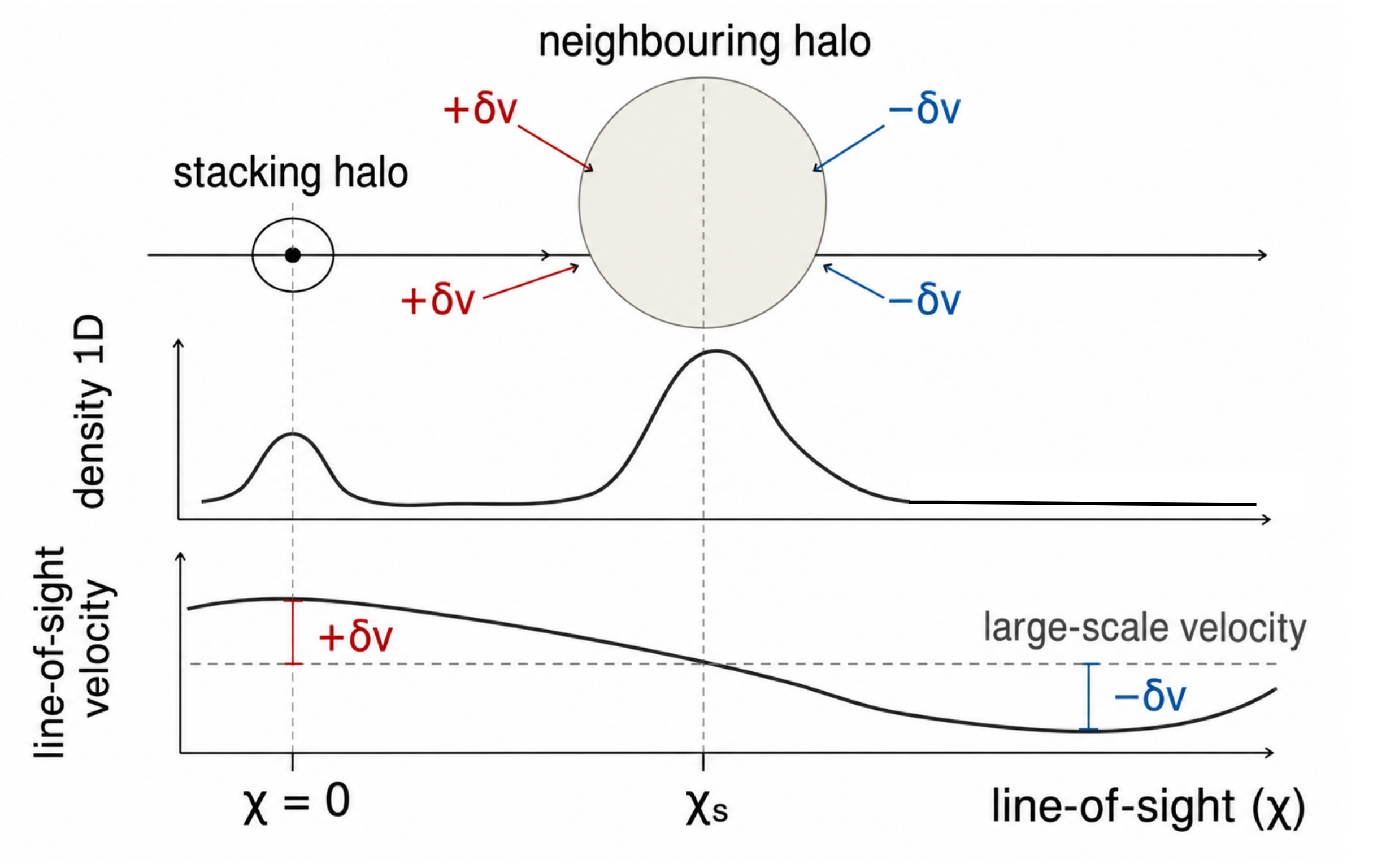} 
    \caption{Schematic diagram illustrating the physical origin of the non-linear velocity terms in the velocity-weighted stacked kSZ signal. In the presence of massive structures in the environment of a halo, its gravitational potential attracts material towards it. In projection, this process creates a positive correlation between projected density and halo velocities, as well as local decorrelation of velocities along the line-of-sight from the halo velocity at the origin.}
    \label{fig:diagram}
\end{figure}

\subsubsection{Physical origin of non-linear terms}\label{sec:non_linear_physics}

The diagram in Figure~\ref{fig:diagram} illustrates the physical process that sources the non-linear terms in the kSZ signal. 
Consider a halo located at $\chi=0$, moving in the positive line-of-sight direction with velocity $v_h$. The halo is embedded within a larger-scale environment rather than evolving in isolation. A nearby massive structure in front of the halo at $\chi_s>0$ gravitationally attracts both the halo and the surrounding gas, producing two distinct effects.

First, the neighbouring structure boosts the halo velocity. Schematically,
\[
v_h \sim v(\chi=0) \sim v_{\rm large\text{-}scale}+\delta v(\delta),
\]
so haloes embedded in overdense environments move faster than average. 

Second, the same neighbouring structure accelerates gas lying beyond it ($\chi>\chi_s$), reducing the gas velocity relative to the halo:
\[
v(\chi>\chi_s)\sim v_{\rm large\text{-}scale}-\delta v(\delta).
\]
The gas therefore becomes progressively decorrelated from the halo bulk motion along the line of sight in the direction of motion of the stacking halo. 


Figure~\ref{fig:correlation_halovel_dens} provides direct evidence for this interpretation. It shows the correlation coefficient between the stacking velocity and either the gas density (upper row) or the gas velocity minus the bulk flow (lower row) in a $20\,h^{-1}\mathrm{Mpc}\times10\,h^{-1}\mathrm{Mpc}$ region around each host halo, where red denotes correlation, blue anti-correlation, and white decorrelation. Stacking haloes are centred on the origin and rotated so that they all move in the positive $\chi$ direction. Statistically, massive neighbouring structures lie in front of the stacking halo, meaning in its direction of motion. The left column uses the true halo velocity, while the middle and right columns use reconstructed velocities discussed in the following sections.

\paragraph*{Density--bulk flow correlation} (upper left panel).
Halo velocities correlate positively with overdensities located in the direction of motion, reaching correlation coefficients of approximately $0.2$. The correlation peaks at projected separations of $2-3\,h^{-1}\mathrm{Mpc}$ and is slightly displaced from the line of sight because perfectly aligned neighbouring structures are statistically rare.

\paragraph*{Velocity decorrelation} (lower left panel).
Gas velocities decorrelate from the halo velocity more rapidly in front of the stacking halo ($\chi>0$) than behind it ($\chi <0$). This asymmetry is precisely what the schematic picture predicts: the neighbouring structure simultaneously accelerates the stacking halo at $\chi=0$ while pulling the gas at $\chi>\chi_s>0$ in the opposite direction relative to the stacking halo, reducing the local coherence of the velocity field.

\subsubsection{Centrals and satellites}\label{sec:cen_sat}

\begin{figure*}
    \centering
    \includegraphics[width=\linewidth]{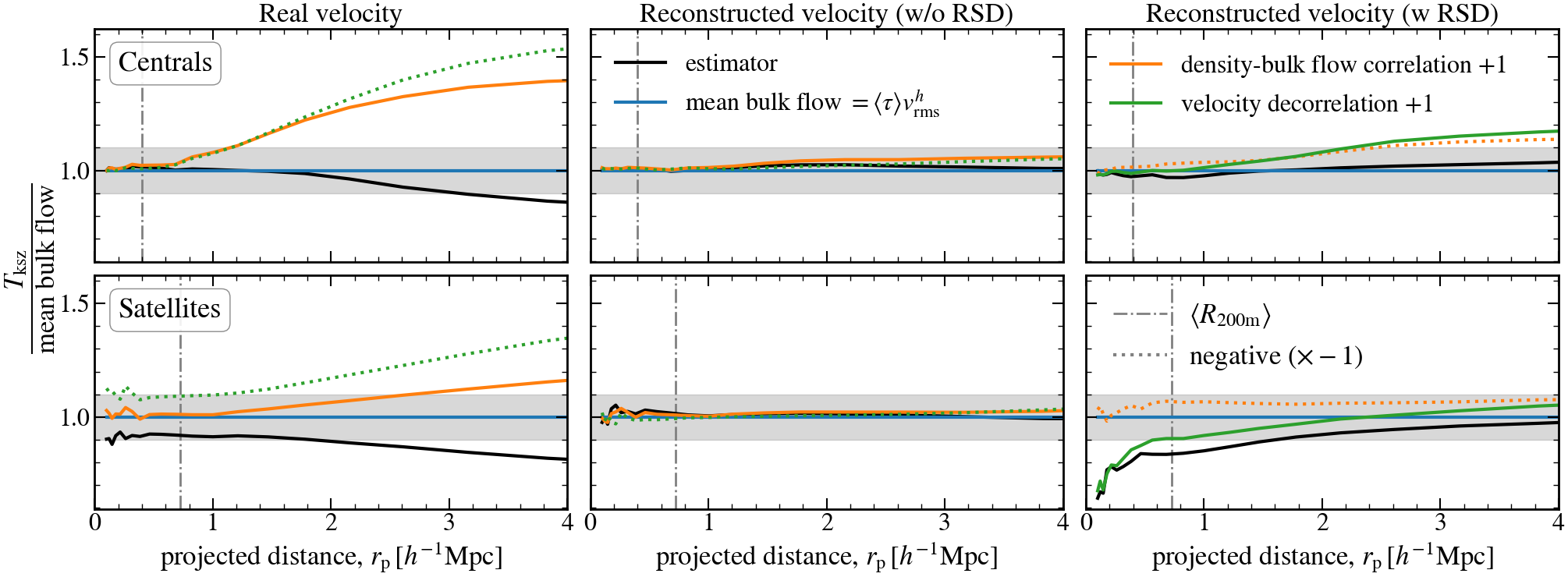}
    \caption{Velocity-weighted stacked kSZ signal around central (upper row) and satellite (lower row) galaxies in the DESI-like LRGs, normalized by the mean bulk flow contribution of the gas -- $T_{\rm ksz} / \langle \tau \rangle v_{\rm rms}^h$. Columns show results for stacking with true halo velocity (left), the reconstructed velocity in real space (middle), and the reconstructed velocity in redshift space including RSD (right). The total signal is shown in black, the density-bulk flow correlation term in orange, the velocity decorrelation term in green, and the mean bulk flow contribution in blue. Dotted curves denote quantities plotted with reversed sign. For clarity, the non-linear terms are plotted as (1+), so that if they are close to 1 it means their contribution to the total signal is negligible. 
    Vertical dash-dotted lines mark the mean virial radius of each sample. Non-linear velocity terms generally suppress the stacked kSZ signal, with the magnitude and scale-dependence depending on galaxy type and stacking velocity. For centrals, the deviation from the mean bulk flow component arises on large scales, whereas for satellites the velocity decorrelation term produces an additional small-scale suppression.}
    \label{fig:ksz_ratios_cen_sat}
\end{figure*}

Projected along the line of sight, these physical processes produce a positive density--bulk flow term and a negative velocity-decorrelation term in the stacked kSZ signal (left column of Figure~\ref{fig:ksz_corr_decorr}). Both effects occur for central and satellite galaxies, but with different relative strengths. The left column of Figure~\ref{fig:ksz_ratios_cen_sat} shows the two populations separately.

\paragraph*{Central galaxies.}

For central galaxies (upper left panel), each non-linear term individually contributes up to $\approx40-50\%$ of the total kSZ signal at projected separations of $\approx4\,h^{-1}\,\rm Mpc$. However, the two terms have nearly identical scale dependence and opposite signs, so they largely cancel. As a result, only a residual non-linear contribution of $\approx10-20\%$ remains on large scales.

This behaviour follows naturally from the physical picture developed in the previous subsection. The same neighbouring structures that correlate the halo velocity with the surrounding density also decorrelate the gas velocity from the halo bulk motion. Since both effects originate from the same gravitational environment, they produce contributions with opposite signs and similar amplitude.

\paragraph*{Satellite galaxies.}

Satellite galaxies show a different balance between the two non-linear terms (lower left panel).

First, the density--bulk flow correlation is weaker, contributing only $\approx20\%$ of the total signal, compared to $\approx40\%$ for centrals. This is expected because satellites reside in more massive host haloes, whose gravitational potential dominates over that of neighbouring structures. Consequently, the surrounding environment has a smaller impact on the bulk velocity of the host halo.

Second, the velocity-decorrelation term becomes substantially larger than the density--bulk flow correlation. It contributes a net $\approx10\%$ correction across nearly all projected scales, remaining significant even close to the satellite position. The origin of this behaviour is the offset of satellite galaxies from the centre of their host halo. Thus, the gas surrounding the satellite no longer shares the bulk velocity of the host halo, contrary to what we have assumed (Section \ref{sec:stack_real_th}). In terms of Eq.~\ref{eq:norm_vel}, this corresponds to $v'(0,0)<0$.

Importantly, this effect is \emph{not} caused by the orbital motions of satellite galaxies. Most of the gas associated with a satellite subhalo is stripped during infall and therefore does not inherit the large orbital velocity of the galaxy \citep{He:2026,Contreras:2026}. Moreover, orbital motions are equally likely to point towards or away from the direction of the halo bulk flow, so their contribution averages to zero in the stacked signal. The same argument applies to mergers: unless the gas velocity is systematically reduced relative to the host-halo velocity, the contribution vanishes after stacking.

Instead, it happens due to the fast, non-linear decorrelation of gas velocity around haloes (as discussed in Section \ref{sec:non_linear_physics}), which is non-negligible even at the distance where satellite galaxies are located. The lower-left panel of Figure~\ref{fig:correlation_halovel_dens} illustrates this effect directly. The farther a satellite lies from the halo centre, the more decorrelated is the gas velocity at that position from the host-halo velocity. For this sample, at distances of $\approx2\,h^{-1}\,\rm Mpc$ from the halo centre, the correlation has already decreased by roughly $50\%$ in the projected direction and by even more along the line of sight. This is another manifestation of non-linear gravity: although linear theory predicts coherent velocities on these scales, this assumption breaks down within the non-linear environments surrounding haloes.

In summary, neighbouring large-scale structures generate the non-linear corrections around both centrals and satellites. For centrals, the two non-linear terms largely cancel because they originate from the same gravitational environment. For satellites, an additional contribution arises because the local decorrelation of gas velocities from the host-halo bulk motion, making the velocity-decorrelation term significant even on small scales. In the next subsections, we show how stacking with a reconstructed velocity modifies this picture.

\subsection{Stacking with reconstructed velocities in real space}\label{sec:stack_vrec_real}

In practice, halo velocities are not directly observable and must instead be reconstructed. We therefore repeat the analysis of Section~\ref{sec:stack_real} using velocities reconstructed from the linearised continuity equation in real space (Section~\ref{sec:stack_vrec_th}). The cross-correlation coefficient of reconstructed and true halo velocities is $r\approx 0.8$.

The result is remarkable: \textit{the non-linear contributions become negligible}. Individually, each contributes less than $10\%$ of the total stacked signal at all projected separations, while their combined contribution remains below $\approx2\%$ (middle panels of Figures~\ref{fig:ksz_corr_decorr} and \ref{fig:ksz_ratios_cen_sat}). The stacked kSZ signal is therefore entirely described by the mean bulk-flow term, for both central and satellite galaxies.

The reason is simple: the physical processes that give rise to the extra terms are intrinsically non-linear, and thus correlate only very weakly with linearly estimated velocities.

This is seen directly in the middle panels of Figure~\ref{fig:correlation_halovel_dens}. The upper-middle panel shows the correlation between the reconstructed velocity and the gas density, with the stacking halo at the origin and moving in the positive $\chi$ direction. As expected from linear theory (Eq.~\ref{eq:vrec}), a weak correlation with the large-scale density field remains at the $<10\%$ level. In contrast, the small-scale signal present when stacking on the true halo velocities (upper-left panel) is almost entirely absent. In other words, the reconstructed velocity is essentially uncorrelated with the small-scale density field around the halo. 

The lower-middle panel shows the correlation between the reconstructed velocity and the deviation of the gas velocity field from the halo velocity. Red regions indicate gas moving coherently with the halo, while white regions indicate gas that has decorrelated from it. Unlike the true halo velocity, the reconstructed velocity is insensitive to the non-linear gravitational processes that both enhance halo velocities and drive the rapid local decorrelation of the surrounding gas. As a consequence, averaging with the reconstructed velocity substantially suppresses the mean gas-halo velocity deviation. A much larger region therefore appears to move coherently with the halo (lower-left vs lower-middle panel).

As a result, both the density--bulk flow (orange) and velocity-decorrelation (green dotted) contributions to the stacked kSZ signal are strongly suppressed (middle panel of Figure~\ref{fig:ksz_corr_decorr}). The same behaviour is seen for both central and satellite galaxies (middle panels of Figure~\ref{fig:ksz_ratios_cen_sat}). Each correction remains below $10\%$, and their combined impact is below $\approx2\%$.

In summary, linear velocity reconstruction in real space filters out the non-linear gravitational processes responsible for both correction terms. The stacked kSZ signal therefore traces the mean bulk flow to within approximately $2\%$ across all scales considered here.

\subsection{Stacking with reconstructed velocities in redshift space}\label{sec:stack_vrec_z}

We now include redshift-space distortions (RSD) in the galaxy positions used for velocity reconstruction. This introduces an additional source of noise in the reconstructed velocity field and further reduces its correlation with the true velocities, to $r \approx 0.7$. 

One might expect RSD to simply reduce the overall amplitude of the stacked kSZ signal through the lower $r$, while leaving its scale dependence unchanged. Yet, this is not the case. The right panel of Figure~\ref{fig:ksz_corr_decorr} shows that the combined non-linear corrections remain negligible on large scales but suppress the signal by $\approx10\%$ on small scales. The individual non-linear terms are smaller than when stacking on the true halo velocities (left panel; Section~\ref{sec:stack_real}), but larger than for the real-space reconstruction (middle panel; Section~\ref{sec:stack_vrec_real}). They also change sign on large scales: the density-bulk flow term becomes negative (orange dotted), while the velocity-decorrelation term becomes positive (green).

This implies that \textit{RSD-induced degradation of the velocity reconstruction correlates with the halo environment and the halo's direction of motion.}

We split the RSD contributions into those induced by the halo environment and those arising from virial motions within the halo itself. The former primarily sources the large-scale features, while the latter dominates on small scales. We examine each in turn below.
\begin{figure*}
 \centering
 \includegraphics[width=\linewidth]{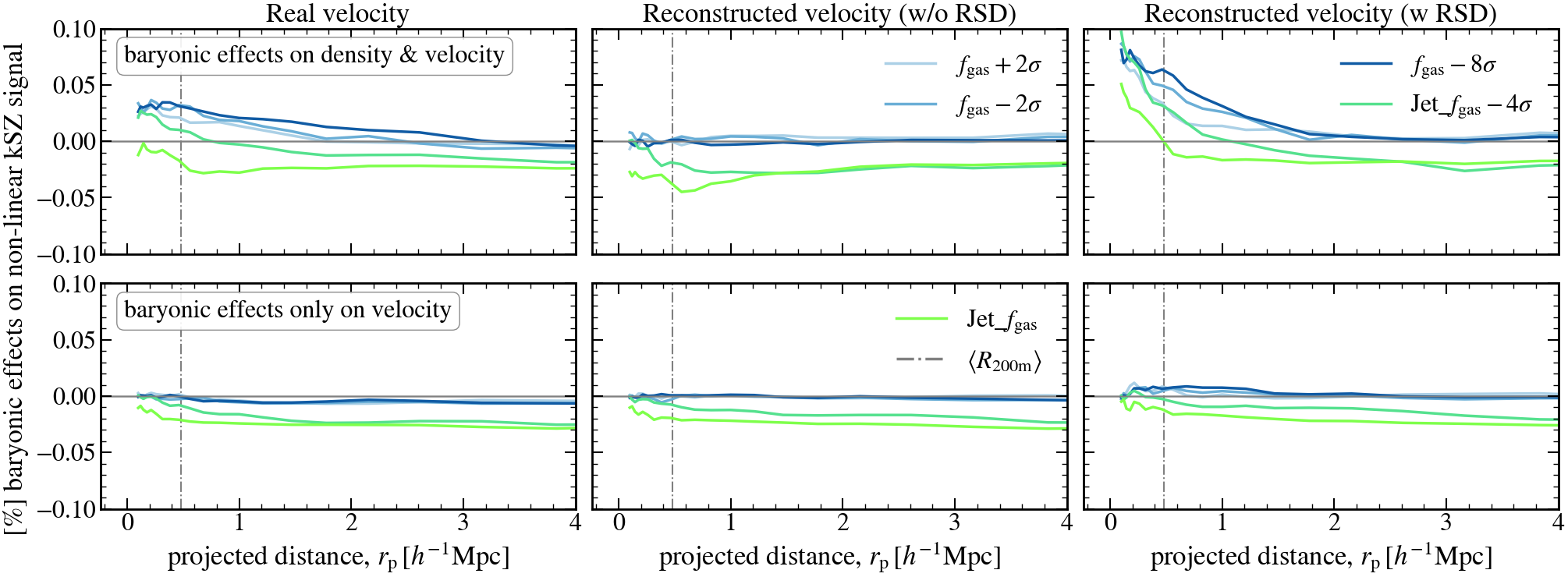}
 \caption{Percentage impact of baryonic physics on the non-linear contribution to the stacked kSZ signal ($T^{\rm nl} = T_{\rm ksz}-\langle\tau \rangle v_{\rm rms }^h$) of DESI-like LRGs,  relative to the mean bulk flow contribution without baryonic effects -- $ (T_{\rm gas}^{\rm nl} - T_{\rm dm}^{\rm nl}) / \langle \tau_{\rm dm}\rangle v_{\rm rms}^h$. The $\rm dm$ underscript denotes the signal if the gas was tracing dark matter distribution and velocities. Zero corresponds to no baryonic impact. Upper panels show the total baryonic effect coming both from gas densities and velocities, while lower panels isolate the contribution from changes in the gas velocity field alone. Columns correspond to different stacking velocities: the true halo velocity (left), reconstructed velocities without redshift-space distortions (RSD; middle), and reconstructed velocities including RSD (right). The vertical dashed line marks the mean halo radius. Colours indicate FLAMINGO feedback variations. Baryonic effects remain below $\approx 5-10\%$ at all radii and for all stacking velocities, with the signal dominated by changes in the gas density distribution rather than by changes in gas velocities.}
  \label{fig:ksz_nonlinear_baryons}
\end{figure*}

\subsubsection{Environment and large-scale features}

The right panels of Figure~\ref{fig:correlation_halovel_dens} show how RSD modify the large-scale behaviour. The non-linear boost in halo velocities from surrounding massive structures (upper left panel, Section~\ref{sec:stack_real}) is mapped into redshift space, producing distortions known as the \textit{Kaiser effect}.

This reduces the performance of the velocity reconstruction in a way that correlates with environmental density: in high-density regions, reconstructed velocities are systematically suppressed, leading to an anti-correlation between reconstructed velocity and environmental density (upper right panel).

The decorrelation of gas velocities from the halo velocity can be understood in a similar way (lower right panel). The reconstructed velocity performs worst in regions dominated by neighbouring massive structures. These are also the regions where most of the gas-halo velocity decorrelation is generated. As a result, averaging with reconstructed velocity reduces the mean gas-halo velocity deviation. The gas therefore appears to move coherently with the halo over a larger region, especially in the direction of halo motion.

After projection, these processes produce a negative large-scale density--bulk flow term (orange dotted line) and a positive large-scale velocity-decorrelation term (green line) in the right panel of Figure~\ref{fig:ksz_corr_decorr}. Combined, the effect is negligible on large scales.

We note that this large-scale behaviour applies to both centrals and satellites, as shown in the rightmost panel of Figure~\ref{fig:ksz_ratios_cen_sat}. The effect is weaker for satellites because they reside in more massive haloes, where the host halo typically dominates the surrounding gravitational potential (Section~\ref{sec:cen_sat}).

\subsubsection{Virial motions and small-scale features}\label{sec:fog}

The virial motions of satellites within their host haloes introduce additional RSD, the so-called \textit{Fingers-of-God} (FoG) effect. As a result, the velocity reconstruction is expected to perform worse in regions dominated by satellites. By itself, this would only reduce the overall signal through lower $r$. A scale-dependent signature requires the FoG distortions to correlate with the halo's direction of motion.

We find that, around satellites, the velocity-decorrelation term suppresses the kSZ signal by up to $\approx30\%$ at the virial radius (lower right panel of Figure~\ref{fig:ksz_ratios_cen_sat}). This translates into a $\approx10\%$ suppression of the total signal, which vanishes on large scales (right panel of Figure~\ref{fig:ksz_corr_decorr}).

A possible explanation is as follows. The halo environment is known to correlate with the internal velocity structure of haloes \citep[e.g.][]{Faltenbacher:2010,Ramakrishnan:2019,Obuljen:2019}. If similar correlations extend to satellite populations, then the strength of FoG distortions would also depend on environment. Since halo environment also correlates with halo motions (Section~\ref{sec:stack_real}), this could naturally generate the residual velocity decorrelation seen in the kSZ stacks.

This effect depends on the smoothing scale used in the velocity reconstruction. The results shown here use $R_s = 10\,h^{-1}\mathrm{Mpc}$. Increasing the smoothing scale reduces the effect, while decreasing it enhances it. For $R_s = 5\,h^{-1}\mathrm{Mpc}$, the suppression exceeds $10\%$ at the virial radius. For $R_s = 30\,h^{-1}\mathrm{Mpc}$, it is about $5\%$. Section~\ref{sec:implications} discusses the impact of the smoothing scale in more detail.


In summary, including RSD in the velocity reconstruction introduces errors that correlate with both the halo environment and the halo's direction of motion. The former generates the large-scale features that cancel, while the latter produces a net suppression of the stacked kSZ signal of $\approx10\%$ on small scales, driven primarily by Fingers-of-God distortions of satellite galaxies.

\subsection{Baryonic effects}\label{sec:non_linear_bar}
In this section, we examine the impact of baryonic feedback on the non-linear contributions to the stacked kSZ estimator. 
We use several feedback variants of the FLAMINGO simulations. In each case, galaxies are selected using the procedure described in Section~\ref{sec:galaxy_selection}. We neglect the back-reaction of baryons on the dark matter, implicitly assuming that the dark matter distribution is the same in the gravity-only and hydrodynamic simulations. Baryonic effects are then defined by comparing the dark matter and gas components within the hydrodynamic simulation.

In all cases, baryonic effects remain below $10\%$ for both the individual and combined non-linear terms. This indicates that these contributions are predominantly gravitational and only weakly affected by baryonic physics.

Figure~\ref{fig:ksz_nonlinear_baryons} quantifies the impact of baryons on the non-linear kSZ contribution,
\begin{equation}
T^{\rm nl}(\rho,v)=T_{\rm ksz}(r_{\rm p})-\langle\tau(r_{\rm p})\rangle v_{\rm rms}^h.
\end{equation}
We compare the gas and dark matter non-linear terms. The dark matter case represents the signal expected if the gas perfectly traced the dark matter density and velocity fields. We define the fractional baryonic effect as
\begin{equation}\label{eq:baryons_dens_vel}
    f^{\rm nl}(\rho, v) =
    \frac{T^{\rm nl}(\rho_{\rm gas},v_{\rm gas})-
    T^{\rm nl}(\rho_{\rm dm},v_{\rm dm})}
    {\langle\tau_{\rm dm}\rangle v_{\rm rms}^h}.
\end{equation}
A non-zero value of $f^{\rm nl}(\rho,v)$ can arise from baryonic effects on the density field, the velocity field, or both. We show the result in the upper panels.

The lower panels isolate the effect of baryons on gas velocities. We assign each gas particle the velocity of its nearest dark matter particle and recompute the non-linear terms. We then define
\begin{equation}\label{eq:baryons_vel}
     f^{\rm nl}(v)=
     \frac{T^{\rm nl}(\rho_{\rm gas},v_{\rm gas})-
     T^{\rm nl}(\rho_{\rm gas},v_{\rm dm})}
     {\langle\tau_{\rm dm}\rangle v_{\rm rms}^h}.
\end{equation}
A non-zero value of $f^{\rm nl}(v)$ therefore measures the contribution from baryonic effects on gas velocities alone. 

\subsubsection{Stacking with true velocities}
The left column shows results for stacks constructed using the true halo velocities. Different colours denote different FLAMINGO variants. Baryonic feedback has only a weak impact on the non-linear terms (upper left panel). Even across substantially different feedback calibrations, the total non-linear contribution remains below $5\%$. We find the same behaviour for the individual density--bulk flow correlation and velocity-decorrelation terms, although each can contribute up to $\approx50\%$ of the total kSZ signal (Section~\ref{sec:stack_real}).

The lower left panel isolates the impact of baryonic effects on gas velocities. For most feedback models, velocity-induced baryonic effects contribute less than $1\%$ to the total kSZ signal. Only the Jet feedback variants reach a few per cent. These models use a different feedback prescription rather than a different calibration\footnote{As pointed out by \cite{OndaroMallea:2025}, the Jet feedback prescriptions of the FLAMINGO simulations exhibit unexpected behaviour even on very large scales, where the velocity power spectra of dark matter and baryons differ by a few per cent. We attribute this to an unexplained numerical effect rather than a physical signal.}.

This weak velocity dependence is remarkable, as baryonic effects on gas velocities remain significant out to scales of $5-10\,R_{\rm 200m}$, depending on the feedback strength \citep{OndaroMallea:2025}. We attribute the weak kSZ response to two effects. First, feedback primarily affects outflowing gas with positive radial velocities, which largely cancel in projection. Second, any residual projected signal contributes to the stacked kSZ measurement only if it is correlated with the stacking velocity, which is not generally expected.

The remaining dependence on gas density should not be interpreted as evidence that baryonic physics generates the non-linear correlations. These correlations arise from non-linear gravitational evolution and are therefore set primarily by the underlying matter distribution, which is dominated by dark matter. Different feedback models simply redistribute the gas within nearly the same matter field. This leads to small changes in the gas-weighted kSZ signal, while leaving the underlying gravitational correlations largely unchanged.

\subsubsection{Stacking with reconstructed velocities}
The middle and right columns show the results obtained using reconstructed velocities in real and redshift space, respectively. As discussed in Section~\ref{sec:stack_vrec_real}, real-space velocity reconstruction strongly suppresses the non-linear terms. As a result, their sensitivity to baryonic feedback is also strongly reduced.

The situation is different in redshift space. In this case, non-linear information leaks into the reconstructed velocity field, making the non-linear contributions to the stacked kSZ signal significant (Section~\ref{sec:stack_vrec_z}). The upper right panel shows that baryonic feedback changes these contributions by $5-10\%$, with the largest effect on small scales. The lower right panel shows that this dependence is almost entirely driven by changes to the gas density field.

We interpret this result in the same way as for stacks constructed with the true halo velocities. The non-linear terms are generated by gravitational dynamics, primarily through the FoG effect of satellite galaxies (Section~\ref{sec:fog}). Baryonic feedback only changes how the gas traces the underlying matter field and its correlations with the velocity fields. 

In summary, baryonic feedback changes the non-linear terms by less than $10\%$. Its contribution through gas velocities alone is below $1\%$. The non-linear terms are therefore predominantly gravitational in origin.
\section{Implications for observation and modelling strategies}\label{sec:implications}
 \begin{figure*}
 \centering
 \includegraphics[width=\linewidth]{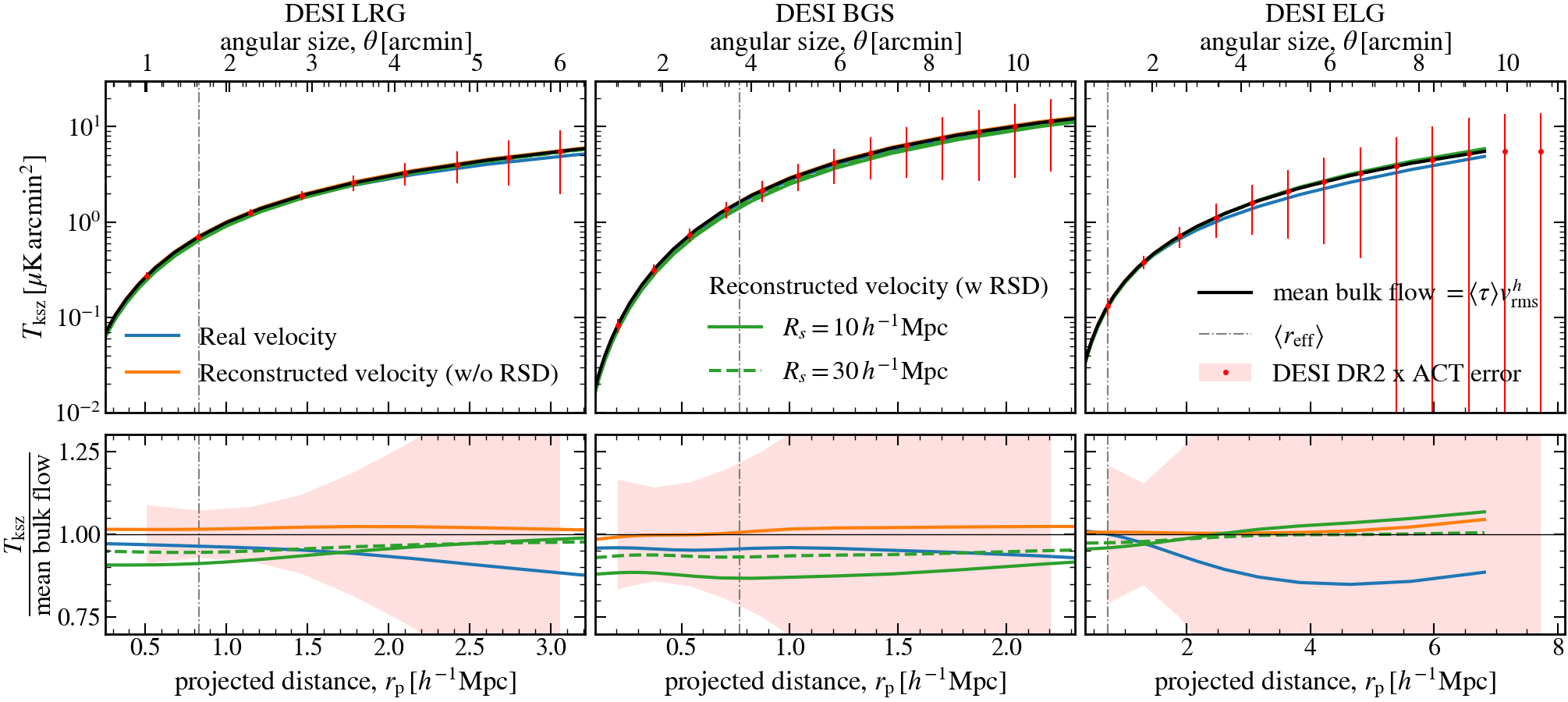}
 \caption{Stacked kSZ profiles obtained using different stacking-velocity estimators for DESI-like LRG (left), BGS (middle), and ELG (right) samples. Upper panels show the predicted signal using the true halo velocity (blue), reconstructed velocity in real space (orange), and reconstructed velocity in redshift space including RSD (green), together with the mean bulk-flow contribution (black). The reconstruction in solid lines uses an aggressive smoothing length of $R_s=10\, h^{-1} \rm Mpc$, while the dashed lines indicate a conservative $R_s=30\, h^{-1} \rm Mpc$. Lower panels show the ratio of each profile to the mean bulk-flow prediction. 
 Vertical dash-dotted lines indicate the effective mean radius of each sample after beam convolution. Red error bars (upper panels) and shaded regions (lower panels) show the relative uncertainty of $\mathrm{ACT} \times \mathrm{DESI \, DR2}$ measurements \citep{Hadzhiyska:2026,Qu:2026}, which employ $R_s=12.5 \, h^{-1}\rm Mpc$ and $R_s=15\, h^{-1}\rm Mpc$ respectively. Different stacking velocity estimators produce scale-dependent differences of approximately $10-20\%$ in the measured signal, which generally suppress it with respect to the mean optical depth of the sample. Increasing the smoothing length in the reconstruction reduces the amplitude of the suppression but does not totally erase it. While the mean bulk flow and measured signals are compatible within $\approx 1-2\sigma$ with DESI DR2 $\times$ ACT measurements, the discrepancy will become statistically significant with the precision targeted by upcoming surveys.}
 \label{fig:scale_dependence_hvel_vrec}
\end{figure*}
The results of the previous section have a number of implications for observational strategies and theoretical models of the stacked kSZ signal.
We discuss these here, distinguishing between the case in which linearly reconstructed velocities are used for stacking --- as in all current observational analyses --- (Section \ref{sec:implications_linear}) and a prospective scenario in which more accurate velocity estimates become available (Section \ref{sec:implications_nonlinear}). Finally, we discuss several approaches from the literature for analysing stacked kSZ data in the context of the results presented in this work (Section \ref{sec:implications_literature}).

As discussed in detail in Section \ref{sec:results}, the stacking velocity slightly changes the shape of the measured kSZ signal, as different stacking velocities couple (or do not couple) differently to the non-linearities in the small-scale gas momentum field. In Figure~\ref{fig:scale_dependence_hvel_vrec} we compare the measured kSZ signal across three realistic galaxy samples --- DESI LRG-like, ELG-like, and BGS-like --- over the range of scales relevant for current observations. We consider several choices of stacking velocity: true halo velocities (blue) and linearly reconstructed velocities in redshift space (green), smoothing the galaxy field at $R_s=10\, h^{-1}\rm Mpc$ (solid) and $R_s=30\, h^{-1}\rm Mpc$ (dashed). For reference, we also show the linearly reconstructed velocities in real space (orange). Note that this is not observable, since we always observe in redshift space in reality. We also plot their ratio to the mean optical depth of the signal (lower panels), and compare them with the fractional uncertainty of the corresponding measurements in the ACT $\times$ DESI DR2 stacks \citep{Hadzhiyska:2026,Qu:2026}.

In order to make the comparison to the measurement errors as realistic as possible, we have applied the beam of the experiment to all profiles. In this case, we assume a Gaussian smoothing with $\text{FWHM}=1.6 \, \rm arcmin$ for all three samples. 

\subsection{Linear velocity reconstruction}\label{sec:implications_linear}

\subsubsection{With RSD}

In a realistic setup, RSD impact the quality of the velocity reconstruction, which propagates into a scale-dependent effect in the measured kSZ profile. In particular, RSD suppress the signal on small scales (green line).

This suppression is primarily driven by the FoGs of satellite galaxies, its amplitude therefore depends on the satellite population of the galaxy sample. For the BGS and LRG samples, where satellites contribute more to the stacked kSZ signal than in the ELG samples, the amplitude of the effect is most significant. 

Moreover, the suppression in the BGS remains at the $\approx 5\%$ level even at $\theta = 10\,\mathrm{arcmin}$. Note that this angular scale corresponds to a projected comoving separation of $r_{\rm p} \approx 2\,h^{-1}\,\mathrm{Mpc}$, where the LRG sample also still exhibits a similar suppression due to RSD. We have verified that, for projected separations $\gtrsim 3\,h^{-1}\,\mathrm{Mpc}$, the RSD-induced suppression decreases for the BGS sample, and the measured profile converges towards the mean bulk flow.


We quantify the statistical significance of the deviation between the measured kSZ signal and the mean bulk-flow approximation, $\delta = T_{\rm ksz} - \langle \tau \rangle v_{\rm rms}^{h}$, using the covariance matrix $C$ of the DESI DR2 $\times$ ACT measurements \citep{Qu:2026,Hadzhiyska:2026}. The significance is defined as
\begin{equation}
    \sigma = \sqrt{\delta\, C^{-1}\, \delta^{\rm T}}.
\end{equation}
For the velocity reconstruction including RSD with a smoothing scale of $R_s = 10\,h^{-1}\,\mathrm{Mpc}$, the measured signal is fully consistent with the mean bulk-flow approximation for ELGs, with $0.31\sigma$. The significance of the deviation increases to $1.14\sigma$ for BGS galaxies and to $1.66\sigma$ for LRGs. Although these deviations are not statistically significant with current data, we expect them to become substantially more significant with the improved sensitivity of next-generation CMB experiments such as the Simons Observatory \citep{Schiappucci:2025,so:2025}. 

An unbiased inference of the mean optical depth therefore requires one of two strategies: 
\begin{enumerate}
    \item \textit{Theoretical}: To explicitly model the scale dependence induced by RSD. This requires an end-to-end simulation-based framework that jointly models the velocity reconstruction, the non-linear gas momentum field, and their coupling, similar to the approach described in Section \ref{sec:implications_nonlinear}. Since the dominant contribution originates from satellite motions, such a framework will likely require galaxy--halo connection models that accurately preserve subhalo positions and velocities.
    \item \textit{Observational}: To suppress the impact of RSD when making the measurement until it becomes subdominant to the observational uncertainties. In this regime, densities and velocities can be treated independently, enabling a decoupled modelling strategy which we discuss in Section \ref{sec:implications_linear_norsd}.
\end{enumerate}

One way to mitigate RSD is to increase the smoothing scale of the density field used in the velocity reconstruction (Eq. \ref{eq:vrec}). The dashed green curves in Figure \ref{fig:scale_dependence_hvel_vrec} show the results for $R_s = 30 \,h^{-1}\mathrm{Mpc}$. Increasing the smoothing scale reduces the suppression of the small-scale signal from approximately $10\%$ to $5\%$ at the effective virial radius for both the LRG and BGS samples. Note that these two choices bracket the smoothing scales adopted in current observational analyses, namely $R_s = 12.5\,h^{-1}\mathrm{Mpc}$ and $R_s = 15\,h^{-1}\mathrm{Mpc}$ \citep{Hadzhiyska:2026,Qu:2026}.

\begin{figure}
    \centering
    \includegraphics[width=0.95\linewidth]{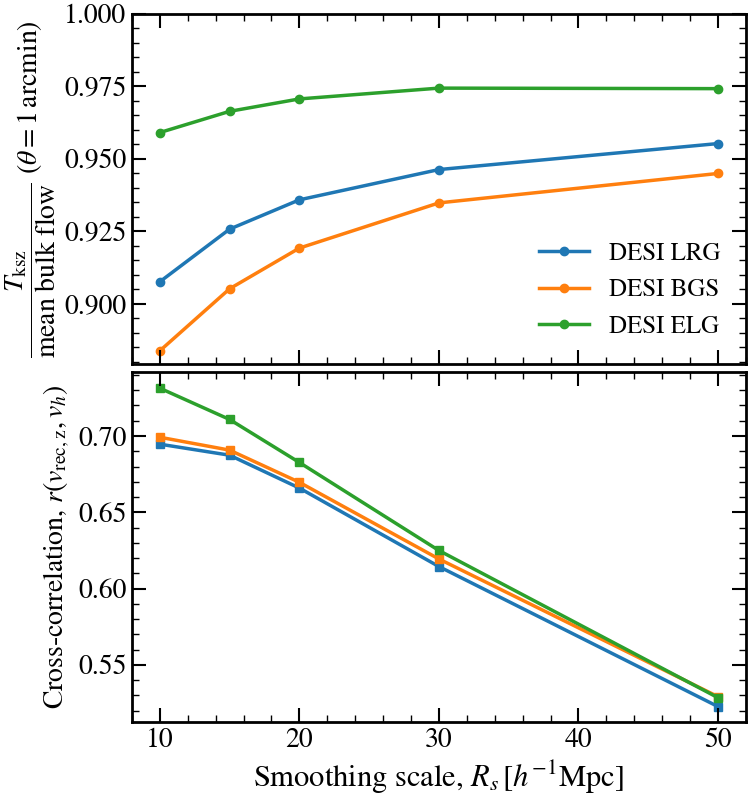}
    \caption{Suppression of the measured kSZ signal relative to the mean bulk flow at $\theta=1\,\mathrm{arcmin}$ using a linear velocity reconstruction in redshift space (upper panel), and the cross-correlation coefficient between the true and reconstructed velocities, $r$ (lower panel), as a function of the reconstruction smoothing scale. Colours correspond to three DESI-like galaxy samples: LRGs (blue), BGS (orange), and ELGs (green). Increasing the smoothing scale reduces the RSD-induced suppression of the measured kSZ signal. For the LRG and BGS samples, the suppression approaches $\approx5\%$ once $R_s \gtrsim 30\,h^{-1}\mathrm{Mpc}$. At the same time, the correlation between the true and reconstructed velocities decreases, reducing the signal-to-noise ratio.}
    \label{fig:vrecz_smoothing}
\end{figure}

This behaviour suggests that the impact of RSD on small scales cannot be suppressed arbitrarily by increasing the smoothing scale. The upper panel of Figure~\ref{fig:vrecz_smoothing} shows the suppression of the measured signal relative to the mean bulk-flow prediction at $\theta = 1\,\mathrm{arcmin}$ as a function of the smoothing scale. As $R_s$ increases from $10\,h^{-1}\,\mathrm{Mpc}$ to $\approx30\,h^{-1}\,\mathrm{Mpc}$, the RSD-induced suppression decreases rapidly. Beyond this scale, however, the suppression saturates at approximately $5\%$ of the mean bulk-flow signal. Increasing the smoothing scale alone therefore cannot reduce the impact of RSD below this level. Using the DESI DR2 $\times$ ACT covariance matrix, this residual suppression at $R_s=30 \, h^{-1} \rm Mpc$ corresponds to a statistical discrepancy of $0.63\sigma$ for the BGS sample and $1.01\sigma$ for the LRG sample. 

The lower panel of Figure \ref{fig:vrecz_smoothing} shows the cross-correlation coefficient between the true and reconstructed velocity fields. As the smoothing scale increases, the correlation coefficient $r$ decreases, leading to a reduction in the signal-to-noise ratio of the kSZ measurement. Increasing the smoothing scale therefore presents a trade-off: it suppresses the impact of RSD and simplifies the modelling, but at the cost of reducing the statistical significance of the measurement.

\subsubsection{Negligible RSD}\label{sec:implications_linear_norsd}

If the impact of RSD is negligible compared with the measurement precision, the measured kSZ signal is well described by
\begin{equation}\label{eq:tksz_rec_lin}
T_{\rm ksz}(r_{\rm p}) \approx \langle \tau (r_{\rm p})\rangle \, v_{\rm rms}^{h}.
\end{equation}

The modelling task therefore reduces to accurately predicting the mean optical depth profile of the galaxy sample as a function of scale, together with the RMS halo velocity and the cross-correlation coefficient between the halo and reconstructed velocities. In this limit, \textit{the velocity field affects only the overall amplitude of the signal} and does not introduce any scale dependence.

Because the gas density is effectively independent of the bulk velocity field, the mean optical depth and velocity statistics can be modelled separately. The optical depth can be calibrated using relatively small hydrodynamic simulations, while the velocity statistics are obtained from much larger gravity-only simulations. This is necessary because the velocity statistics are much more sensitive to large-scale modes than the optical depth and therefore require much larger simulation volumes to converge.

If the cross-correlation coefficient or the RMS halo velocity, $r$, is poorly constrained, a conservative approach is to marginalize over the overall signal amplitude. Although this reduces the constraining power of the measurement, it also makes the analysis more robust to uncertainties in the velocity modelling. In this case, the information contained in the amplitude of the optical depth profile is discarded, while the information encoded in its scale dependence is retained. Quantifying the relative constraining power of the amplitude and the shape of the optical depth profile is beyond the scope of this work.

\begin{table*}
\centering
\caption{Summary of the main conclusions of the paper. Note that the linear reconstruction without redshift-space distortions does not exist in practice, as all observations are done in redshift space. In all regimes the non-linear terms are gravitational in origin and only
weakly sensitive to baryonic feedback ($<10\%$; baryonic effects on gas velocities
alone contribute $<1\%$), so gravity-only simulations with baryon-painting of the
density field suffice. At the current DESI DR2 $\times$ ACT precision, the statistical significance of non-linear terms is of $\approx 1-2\sigma$ in the measured signal for linearly reconstructed velocities including RSD with $R_s = 10\,h^{-1}\,\mathrm{Mpc}$, decreasing to $\lesssim 1\sigma$ for $R_s = 30\,h^{-1}\,\mathrm{Mpc}$.}
\begin{tabularx}{\textwidth}{@{}p{2.4cm} X X X X X@{}}
\toprule
\textbf{Velocity estimator} & \textbf{Smoothing scale} & \textbf{S/N} & \textbf{What the signal traces} & \textbf{Non-linear terms} & \textbf{Modelling required} \\
\midrule
Linear reconstruction, no RSD
  & $10\,h^{-1}\rm Mpc$
  & -- (not observable)
  & Mean optical depth $\times$ overall velocity amplitude
  & Negligible ($\approx 2\%$)
  & \textit{Linear regime:} Decoupled densities $\&$ velocities: $\tau$ from small boxes, velocity statistics from large gravity-only boxes \\
\addlinespace
\addlinespace
\multirow{2}{=}{Linear reconstruction, with RSD}
  & $>30\,h^{-1}\rm Mpc$
  & Lower; $r<0.7$
  & \multirow{2}{=}{Mean optical depth $+$ scale-dependent suppression (satellite FoG--driven)}
  & Up to $\approx 5\%$ (small scales)
  & \multirow{2}{=}{Depending on the precision of the measurement, \textit{linear} or \textit{non-linear regime}} \\
  & $10\,h^{-1}\rm Mpc$
  & Intermediate; $r\approx 0.7$
  &
  & Up to ${\approx}10\%$ (small scales)
  & \\
  \addlinespace
\addlinespace
Beyond-linear (non-linear / ML)
  & --
  & Higher; $r>0.7$
  & Mean optical depth $+$ scale-dependent suppression
  & Up to $\approx 10-20\%$
  & \textit{Non-linear regime:} Joint, simulation-based: gas momentum field $+$ velocity reconstruction $+$ their coupling; the two non-linear terms must be modelled \textit{together} \\
\bottomrule
\end{tabularx}
\label{tab:ksz_regimes}
\end{table*}

\subsection{Beyond linear reconstruction}\label{sec:implications_nonlinear}
 
If future velocity reconstruction algorithms move beyond linear theory---for example through non-linear \citep{Kitaura:2012} or machine-learning methods \citep{Tanimura:2022,Veena:2023,Tanimura:2024,Chen:2024,Wang:2024,Troster:2026,Gong:2026}---the non-linear contributions identified in this work will re-emerge. They will modify the stacked kSZ signal by $10-20\%$ in a scale-dependent way. This is illustrated by the blue curves in Figure~\ref{fig:scale_dependence_hvel_vrec} for the three galaxy samples. Modelling this regime requires two additional ingredients: the correlation between bulk gas flows and the surrounding small-scale density, and the local decorrelation of gas velocities from the bulk flow.

These two effects must be modelled jointly. As shown in Section~\ref{sec:stack_real}, they arise from the same underlying physical process and have opposite signs. Modelling only one of them can produce a larger systematic error than neglecting both. For example, including the density--bulk flow correlation while neglecting velocity decorrelation ($T_{\rm ksz}\propto\langle\tau v_h^2\rangle$) leads to errors of $\approx50\%$ at $r_{\rm p}\approx4\,h^{-1}\mathrm{Mpc}$. Neglecting both effects ($T_{\rm ksz}\propto\langle\tau\rangle\langle v_h^2\rangle$) reduces the error to only $\approx10-20\%$.

The non-linear terms are primarily generated by gravitational dynamics and depend only weakly on baryonic physics (Section~\ref{sec:non_linear_bar}). They can therefore be modelled to the $\approx5\%$ level using gravity-only simulations. The remaining baryonic contribution is dominated by changes to the gas density field, while baryonic effects on gas velocities contribute less than $1\%$ to the stacked kSZ signal. Current baryon-painting approaches, which assign gas densities to gravity-only simulations while leaving the velocity field unchanged \citep{Osato:2023,Arico:2024,Anbajagane:2024,Schneider:2025}, therefore remain valid. Extending these methods to model baryonic effects on gas velocities is unnecessary.

\subsection{Implications for existing kSZ analyses}\label{sec:implications_literature}

In this section, we review several recent theoretical studies of the stacked kSZ signal measurements and discuss the implications of our results for their modelling strategies.

The analysed data consist of stacks of ACT CMB maps at galaxy positions, weighted by linearly reconstructed velocities, using either BOSS galaxies \citep{Schaan:2021} or DESI galaxies \citep{Hadzhiyska:2024,RiedGuachalla:2025,Qu:2026,Hadzhiyska:2026}.

Broadly, two approaches have been adopted in the literature. The first models the kSZ signal directly using hydrodynamic simulations \citep{McCarthy:2025b,Bigwood:2025,Siegel:2025}, while the second relies on analytic descriptions \citep{Bigwood:2024,Kovac:2025,Siegel:2026}. A common feature of both approaches is that they only consider the impact of velocity reconstruction through the cross-correlation coefficient $r$, which rescales the overall signal amplitude and must be accurately characterized when fitting for the kSZ amplitude. Here, we go one step further by focusing on the scale-dependent effects arising from non-linear velocities and their coupling to the stacking velocity.

In hydrodynamic simulations, the simulated kSZ signal naturally includes all non-linear information, including small-scale density--velocity correlations. Existing works then construct the stacked estimator using the true \textit{galaxy} velocities\footnote{In Appendix~\ref{sec:app_stack_vel}, we show that using the velocities of central objects---whether those of host haloes, central subhaloes, or the stars associated with the central halo \citep[e.g.][]{Bigwood:2025,Siegel:2025}---yields virtually identical profiles. Therefore, the conclusions presented in this work remain unchanged. By contrast, when galaxy velocities, including those of satellite galaxies, are used \citep[e.g.][]{McCarthy:2025b}, the scale-dependent effects and deviations from the mean bulk flow become more pronounced on large scales. Thus, although the qualitative behaviour remains the same, the quantitative impact of non-linear effects is larger than what we discuss in this section.} from the simulation \citep{McCarthy:2025b,Bigwood:2025,Siegel:2025}. This prediction is then compared with measurements based on linearly reconstructed velocities. Although both the theoretical prediction and the measurements include the mean bulk flow and non-linear terms, the latter depend on the choice of stacking velocity (Section~\ref{sec:results}). This is equivalent to comparing the blue (theory) and green (measurement) curves in Figure~\ref{fig:scale_dependence_hvel_vrec}, which differ by up to $5-10\%$ in a scale-dependent manner despite having identical galaxy selection and gas distribution. This demonstrates the importance of treating the stacking velocity \textit{consistently} in both theory and data.

An alternative is to use analytic models to predict the stacked kSZ signal \citep{Bigwood:2024,Kovac:2025,Siegel:2026}. These models describe only the mean bulk flow (through the
mean optical depth) and neglect the non-linear contributions. Comparing such predictions with measurements is therefore equivalent to comparing the black and green curves in Figure~\ref{fig:scale_dependence_hvel_vrec}. Again, scale-dependent differences of approximately $5-10\%$ arise, although they can be reduced by adopting a larger velocity reconstruction smoothing scale, $R_s \geq 30\,h^{-1}\mathrm{Mpc}$.\footnote{Existing models of the kSZ effect are inspired by \textit{baryonification} techniques but are restricted to analytic gas profiles. Recent developments in baryonification techniques \citep[e.g.][]{Arico:2024,Schneider:2025}, however, make it possible to construct simulation-based baryonification models for the kSZ effect that inherit the full non-linear density and velocity fields, similarly to hydrodynamic simulations.}

Furthermore, \cite{Siegel:2026} quantify the impact of velocity decorrelation in their Appendix D1 using FLAMINGO simulations. They compare a kSZ signal computed with the true gas particle velocities to one in which the gas is assumed to move with the velocity of the central galaxy, finding differences of approximately $50\%$ on scales of a few arcminutes. We caution that this comparison may overestimate the physical impact of velocity decorrelation. As discussed in Section~\ref{sec:implications_nonlinear}, it effectively assumes
\begin{equation}
T_{\rm kSZ} \propto \langle \tau v_h^2\rangle
= \langle\tau\rangle\langle v_h^2\rangle
+ \mathrm{Cov}(\tau,v_h^2),
\end{equation}
which is a much poorer approximation to the true kSZ signal than retaining only the first term -- mean bulk flow. This again highlights the need to model the non-linear contributions jointly: including only one effect (here, the correlation between halo velocity and density) while neglecting the other (velocity decorrelation) produces a much larger error ($\approx 50\%$) than neglecting both ($\approx 10-20\%$), because both originate from the same physical process and largely compensate each other (see Section \ref{sec:stack_real}).

At the current observational precision, inconsistencies in the treatment of non-linear velocities and velocity reconstruction between theoretical models and measurements are not expected to produce statistically significant biases (typically at the $1-2\sigma$ level). Consequently, the conclusions of the studies discussed above are likely not affected by these inconsistencies. However, the next generation of stacked kSZ measurements is expected to achieve substantially higher precision, making these effects increasingly important. Theoretical modelling will therefore need to account for them consistently.

\section{Summary and conclusions}\label{sec:conclusions}

The velocity-weighted stacked kSZ signal is emerging as a powerful probe of the baryon distribution around galaxies, but its interpretation relies on assumptions about the underlying velocity field that have not previously been tested systematically. In this work, we use the FLAMINGO cosmological hydrodynamical simulations together with realistic DESI-like galaxy mocks (LRGs, ELGs, and BGS galaxies) to study how non-linear velocity effects contribute to the stacked kSZ signal and to assess their impact on current modelling approaches. We summarise our conclusions in Table  \ref{tab:ksz_regimes}.

Decomposing the signal into a mean bulk flow component and non-linear velocity contributions (Table \ref{tab:ksz_decomposition}), we find that the non-linear terms can account for $10-20\%$ of the total signal. These contributions are generated by non-linear gravitational dynamics that significantly affect gas velocities on small scales, inducing correlations with the density field and reducing local velocity coherence. Whether these effects appear in the stacked kSZ signal at all, and how they depend on scale, is determined by their coupling with the stacking velocity.

When true halo velocities are used for stacking, these non-linear contributions account for $10-20\%$ of the measured kSZ signal on large scales (Section~\ref{sec:stack_real}). The common assumption that velocities remain constant and independent of the small-scale density distribution across the scales probed by the stacked kSZ signal is therefore not justified in general.

In practice, halo velocities must be estimated, typically through linear reconstruction methods based on the continuity equation. If the reconstruction is performed in real space, which is not possible in reality, the non-linear processes affecting the gas velocity field do not correlate with the linearly reconstructed stacking velocities. As a result, the non-linear contributions are effectively washed out, and the stacked kSZ signal is accurately described by the mean optical depth profile multiplied by a single velocity-dependent normalization factor (Section~\ref{sec:stack_vrec_real}).

The final step towards realism is to include redshift-space distortions (RSD) in the velocity reconstruction. In this case, the reconstruction performance becomes correlated with the halo environment, introducing scale-dependent modifications to the measured kSZ profile and suppressing its amplitude on small scales by $\approx 10\%$ (Section~\ref{sec:stack_vrec_z}). The impact of RSD on small scales can be reduced to $\approx 5\%$ by increasing the smoothing scale in the velocity reconstruction to $R_s \approx 30 \, h^{-1}\rm Mpc$. However, this also reduces the cross-correlation coefficient between true and reconstructed velocities, and hence the signal-to-noise ratio.

A key result of this work is that the non-linear velocity contributions are remarkably insensitive to baryonic feedback, which change the profile by $<5-10\%$, since they are generated by gravitational non-linearities (Section~\ref{sec:non_linear_bar}). Notably, the baryonic effects on the gas velocities alone impact the kSZ signal by $<1\%$.

These findings have important implications for the interpretation of stacked kSZ measurements (Section~\ref{sec:implications}). When the signal is measured with linearly reconstructed velocities and negligible redshift-space distortion effects, the measured signal directly traces the mean optical depth of the galaxy sample, with the velocity statistics only affecting the overall amplitude of the signal. In this limit, the gas distribution and velocity statistics can be modelled independently, greatly simplifying theoretical predictions. This is the regime of current data within $\approx 1-2 \sigma$. If, however, future reconstruction methods recover non-linear information or are affected by redshift-space distortions, the scale-dependent non-linear gravitational effects on velocities will need to be modelled, likely requiring fully simulation-based approaches that also incorporate the velocity reconstruction.

More generally, our results demonstrate that consistency between the observational estimator and the theoretical model is essential. Stacking with true versus reconstructed velocities produces scale-dependent differences of up to $10-20\%$, implying that mismatched modelling assumptions can induce biases of order $\approx 1-2 \sigma$ already at the current statistical precision of the data, and exceeding several $\sigma$ for upcoming measurements \citep{so:2025}.

The principal trade-off is therefore between \textit{statistical precision and modelling complexity}. Linear velocity reconstruction remains robust and simple to interpret, even in redshift space, but only for large smoothing scales of $R_s \approx 30\,h^{-1}\mathrm{Mpc}$. Reducing the smoothing scale, or adopting more sophisticated velocity reconstruction methods, increases the statistical power of the measurement at the cost of introducing non-linear effects that are already comparable to current uncertainties and will require accurate modelling for next-generation kSZ measurements.

\section*{Acknowledgements}
REA received support from grant PID2024-161003NB-I00 funded by MICIU/AEI/10.13039/501100011033 and by ERDF/EU. \textit{Author contributions}: \textbf{LOM}: conceptualization; formal analysis; methodology; investigation; visualization; validation; writing-original draft; \textbf{REA}: conceptualization, formal analysis, writing-review\&editing; \textbf{BH}: formal analysis, writing-review\&editing; \textbf{JS}: project administration (FLAMINGO); data curation; resources; writing-review\&editing. 

\appendix
\section{Choice of stacking velocity in simulations}\label{sec:app_stack_vel}

\begin{figure}
    \centering
    \includegraphics[width=0.8\linewidth]{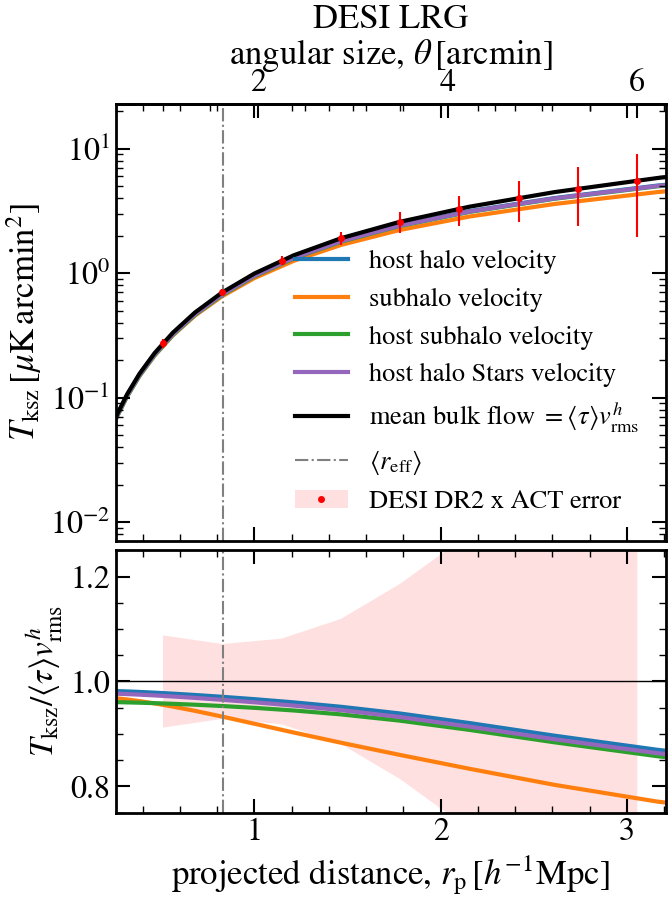}
    \caption{Comparison of the stacked kSZ profiles of the DESI LRG-like sample for different choices of stacking velocity in the simulations. The profiles are obtained using the host halo (blue), satellite subhalo (orange), host (central) subhalo (green), and host halo stellar (purple) centre-of-mass velocities. The black curve shows the mean bulk flow component, assuming the rms host halo velocity. The upper panel shows the stacked profiles, while the lower panel shows their ratio to the mean bulk flow. The red points and shaded band indicate the statistical uncertainties expected from the DESI DR2 $\times$ ACT measurements \citep{Qu:2026}. Velocity definitions based on the host halo or central subhalo produce nearly identical profiles, whereas using satellite subhalo velocities introduces a stronger scale dependence.}
    \label{fig:compare_stacking_vel}
\end{figure}

The choice of stacking velocity in simulations is not straightforward. Figure \ref{fig:compare_stacking_vel} compares the profiles obtained for the DESI LRG-like sample using several stacking velocities. The black curve shows the mean bulk flow component, assuming the rms host halo velocity, while the red curve indicates the relative statistical uncertainty expected from the DESI DR2 $\times$ ACT measurements of \citet{Qu:2026}. The upper panel presents the profiles, and the lower panel their ratio to the mean bulk flow.

Our fiducial choice is the host halo centre-of-mass velocity (blue). As discussed in the main text (Section \ref{sec:stack_real}), this results in a suppression of approximately $5\%$ relative to the mean bulk flow on small angular scales, increasing to $\approx15\%$ at $\theta = 6\,\rm arcmin$. Alternative velocity definitions based on the host halo or the central subhalo, such as the central subhalo velocity (green) and the stellar centre-of-mass velocity of the host halo (purple), produce very similar profiles.\footnote{We do not show the profiles obtained using the halo gas centre-of-mass velocity because of a numerical constraint: gas centre-of-mass velocities are computed only for haloes containing more than 100 bound gas particles. However, we have verified that, when restricting the analysis to haloes unaffected by this constraint, the resulting kSZ profiles are equivalent to those obtained using the halo centre-of-mass velocity.} This agreement is expected, but nevertheless reassuring, as it indicates that the large-scale bulk flow is largely independent of the particle species used to estimate it. 

The largest differences arise when the stacking is performed using the velocities of satellite subhaloes instead of those of their host haloes. This case is shown by the orange curve in Figure \ref{fig:compare_stacking_vel}. On small scales, the suppression relative to the mean bulk flow is comparable to that of the other velocity definitions. However, the suppression becomes significantly stronger at larger angular scales, reaching approximately $20\%$ at $\theta = 6\,\rm arcmin$.

These results motivate our choice of host halo or central velocities over satellite subhalo velocities when constructing stacked kSZ profiles in simulations. First, the resulting profiles exhibit weaker scale-dependent velocity effects and therefore provide a closer approximation to the mean bulk flow. Second, satellite velocities are observationally much more difficult to estimate because of their highly non-linear dynamics \citep[e.g.][]{RiedGuachalla:2023}, making a direct comparison between theoretical predictions and observational measurements considerably more challenging.

\bibliographystyle{mnras}
\bibliography{main}

\bsp	
\label{lastpage}
\end{document}